\documentclass[review]{elsarticle}

\usepackage{hyperref}

\journal{Journal of Computational Physics}

\usepackage[utf8]{inputenc}
\usepackage[fleqn]{amsmath}
\usepackage{amsfonts,amsthm}
\usepackage{amssymb,graphicx,subfigure,color,}
\allowdisplaybreaks[1]

\numberwithin{equation}{section}
\numberwithin{figure}{section}
\numberwithin{table}{section}

\newcommand{\ds}{\displaystyle}
\newcommand{\rp}{\rho^{n+1}}
\newcommand{\rn}{\rho^{n}}

\newcommand{\un}{{\bf v}^{n}}
\newcommand{\rup}{\rho^{n+1} {\bf v}^{n+1}}
\newcommand{\run}{\rho^n {\bf v}^{n}}
\newcommand{\cp}{c^{n+1}}

\newcommand{\dt}{\Delta t}
\newcommand{\dx}{\Delta x}
\newcommand{\dy}{\Delta y}
\newcommand{\eps}{\varepsilon}
\newcommand{\rf}[1]{(\ref{#1})}
\newcommand{\vecu}{{\bf u}}

\newcommand{\vecv}{{\bf v}}
\newcommand{\vecn}{{\bf n}}
\newcommand{\vecf}{{\bf f}}

\newcommand{\vecx}{{\bf x}}

\newcommand{\T}{{\bf T}}
\newcommand{\matD}{{\bf D}}
\def\setR {\mathbb{R}}

\def\matH {{\bf H}}
\newcommand{\ijp}{_{i,j}^{n+1}}
\newcommand{\ijn}{_{i,j}^{n}}
\newcommand{\pjp}{_{i+1,j}^{n+1}}
\newcommand{\pjn}{_{i+1,j}^{n}}
\newcommand{\mjp}{_{i-1,j}^{n+1}}
\newcommand{\mjn}{_{i-1,j}^{n}}
\newcommand{\ipp}{_{i,j+1}^{n+1}}
\newcommand{\ipn}{_{i,j+1}^{n}}
\newcommand{\imp}{_{i,j-1}^{n+1}}
\newcommand{\imn}{_{i,j-1}^{n}}
\newcommand{\uz}{^{(0)}}
\newcommand{\uo}{^{(1)}}
\newcommand{\ut}{^{(2)}}
\newcommand{\pz}{_{i,j}^{n+1,(0)}}
\newcommand{\nz}{_{i,j}^{n,(0)}}

\newcommand{\po}{_{i,j}^{n+1,(1)}}
\newcommand{\no}{_{i,j}^{n,(1)}}

\newcommand{\pt}{_{i,j}^{n+1,(2)}}
\newcommand{\nt}{_{i,j}^{n,(2)}}

\theoremstyle{definition}
\newtheorem{thm}{Theorem}[section]

\newtheorem{exa}[thm]{Example}

\begin{document}

\begin{frontmatter}

\title{An Asymptotic-Preserving Method for a Relaxation of the Navier-Stokes-Korteweg Equations}
\author[acaddress]{Alina Chertockd}
\ead{chertock@math.ncsu.edu}

\author[pdaddress]{Pierre Degond}
\ead{p.degond@imperial.ac.ukm}

\author[jnaddress]{Jochen Neusser\corref{mycorrespondingauthor}}
\cortext[mycorrespondingauthor]{Corresponding author}
\ead{jochen.neusser@mathematik.uni-stuttgart.de}

\address[acaddress]{Department of Mathematics, North Carolina State University State University, Raleigh, NC, 27695, USA}
\address[pdaddress]{Department of Mathematics, Imperial College London, London SW7 2AZ, UK}
\address[jnaddress]{Universit\"at Stuttgart, Institut f\"ur Angewandte Analysis und Numerische Simulation, Pfaffenwaldring 57,D-70569 Stuttgart, Germany}

\begin{abstract}
The Navier-Stokes-Korteweg (NSK) equations are a classical diffuse-interface model for compressible two-phase flow. As direct numerical simulations based  on the NSK system are quite expensive and in some cases even impossible, we consider a relaxation of the NSK system, for which robust numerical methods can be designed.
However, time steps for explicit numerical schemes depend on the relaxation parameter and therefore numerical simulations in the relaxation limit are very inefficient.
To overcome this restriction, we propose an implicit-explicit asymptotic-preserving finite volume method. We prove that the new scheme provides a consistent discretization of the NSK system in the relaxation limit and demonstrate that it is capable of accurately and efficiently computing numerical solutions of problems with realistic density ratios and small interfacial widths.
\end{abstract}


\end{frontmatter}

\section{Introduction}
\label{sec:intro}
There are in general two approaches to describe the behavior of multi-phase fluids, the sharp interface (SI) and the diffuse interface (DI) approach. The first approach represents multiple phases with different sets of equations that are coupled by some interface conditions. The second approach, which used to describe, e.g. the merging process of droplets and bubbles, needs only one set of equations to model the phases and does not require the location of the interface to be tracked explicitly.

In this paper, we consider two DI models for a homogeneous two-phase compressible fluid:  The Navier-Stokes-Korteweg system (NSK) and a relaxation system for the NSK system. The NSK system goes back to the work of Korteweg \cite{Korteweg-01} and was formulated in its present form in \cite{DunnSerrin-85,AMW-98}. The NSK system uses a Van-der-Waals like pressure function to identify two distinct phases and a third-order term to model phase transitions. Many authors achieved analytical results on the well-posedness of the NSK system and its variants, e.g. \cite{Benzoni99,Bresch-03,HattoriLi-94,Kotschote-08,Rohde-10}. While some numerical methods have been successfully developed and used for these and related families of problems, see, e.g., \cite{BraackProhl},\cite{Charve13},\cite{Diehl},\cite{GLT15},\cite{Hermsdorferetal},\cite{Tadmor-84}, there are still many open problems, for which accurate and efficient numerical methods are yet to be designed. Up to our knowledge, the robust computation of realistic density values for liquid and vapor phases has not been suggested for the NSK model. Additionally, numerical methods also fail in cases when very small interface widths close to a sharp interface are to be considered. In both cases, the occurring problems are related to steep density gradients. Another source of difficulty one comes across while numerically solving NSK systems is related to the Van-der-Waals like form of the pressure equation. The latter prevents one from using upwind hyperbolic solvers, which have been successfully applied, e.g., to stabilize computations for Navier-Stokes equations with high Reynolds numbers.

The issue of very small interfaces is especially important because the NSK model can only provide the correct amount of capillary forces if the interface is extremely small \cite{DreyerGiesselmannKrausRohde,Hermsdorferetal,Jamet-01}. One idea to loosen the strict coupling between interfacial width and capillary forces is to introduce an additional Cahn-Hilliard or Allen-Cahn type equation for a new phase field variable. This was done for example in \cite{ADGK13,Blesgen,Witterstein}. 
Another ansatz to avoid some of the difficulties for the NSK systems suggests to introduce a relaxation of the NSK system, in which the third-order term is replaced by a first-order term and a Poisson equation, that defines a new  phase field parameter, see, e.g., \cite{Rohde-10}. This model is parametrized by a so-called Korteweg parameter $\alpha$. If the Korteweg parameter tends to zero, the relaxation system formally converges to the NSK system. The most important feature of the relaxation system is the fact, that the first-order part is purely hyperbolic for sufficiently small Korteweg parameter. It should also be observed that the addition of the Poisson equation to the system does not increase the computational cost of numerical simulations as test cases demonstrate that the time savings, that come from the fact that one does not have to solve a third order system, are greater than the loss of time that comes from the numerical solution of the Poisson equation.  These properties can be exploited to construct robust numerical schemes for the relaxation system.  In \cite{NeusserRohdeSchleper-15}, it has been shown that the overall approach is robust for problems with large density ratios and small interfacial widths. However, the numerical scheme proposed in \cite{NeusserRohdeSchleper-15} is an explicit scheme and thus the time steps decrease as the Korteweg parameter $\alpha$ tends to zero. 

It is the main purpose of this contribution to construct an asymptotic-preserving (AP) scheme \cite{Jin-99} in the Korteweg limit, that is, a scheme for the relaxation system that provides a consistent approximation of the original NSK system as the Korteweg parameter $\alpha$ tends to zero.  The AP approach was developed in the framework of linear transport in diffusive regimes \cite{Jinetal-99,JinLevermore-91,Larsen-87} and has been applied to many different areas, e.g. fluid and diffusion limits of kinetic models, relaxation methods for hyperbolic systems and low-mach number limits for compressible flow problems; see, e.g.,  \cite{CGRS,CDK,DegondJinLiu-07,DT11,Giesselmann-15,Haack-12,Kle95,ANLM,PR03}. In \cite{Haack-12},  the time and spatial discretization of the isentropic Euler and Navier-Stokes Equations in the low Mach number limit was investigated. Inspired by this research, we construct here a scheme that captures the Korteweg limit for the relaxation system and prove the AP property of the proposed scheme. The AP property of the new scheme is achieved by splitting the relaxation system into a non-stiff nonlinear, compressible hyperbolic Navier-Stokes like system and a system that can be treated by a Poisson solver,  and  allows the use of time and spatial steps that are independent of the Korteweg parameter. As the result the proposed numerical scheme is very efficient for small values of the parameter $\alpha$, which is a significant improvement compared to an explicit scheme from \cite{NeusserRohdeSchleper-15}. Beyond that, we expect our scheme to be asymptotic preserving in the sharp interface limit. We cannot give analytical proof to that, but we support this statement by a numerical example. 

The paper is organized as follows. In Section \ref{sec:NSK}, we introduce the NSK and relaxation systems and describe their main properties together with the basic thermodynamical framework. We comment on the advantages of the relaxation system and point out why it is necessary to introduce a new scheme in order to solve the relaxation system efficiently. 
Section \ref{sec:allspeed} contains the basic outcome of this contribution. We propose the AP scheme for the relaxation system and perform an asymptotic analysis to show that the scheme transforms into a scheme for the NSK equations in the Korteweg limit. In Section  \ref{sec:numerics}, we demonstrate that the algorithm provides a massive improvement compared to a standard explicit scheme for a number of problems in one and two space dimensions. 

\section{Navier-Stokes-Korteweg equations}
\label{sec:NSK}
\subsection{The Navier-Stokes-Korteweg system}
\begin{figure}[ht!]
\centering
\includegraphics[width=0.5\columnwidth]{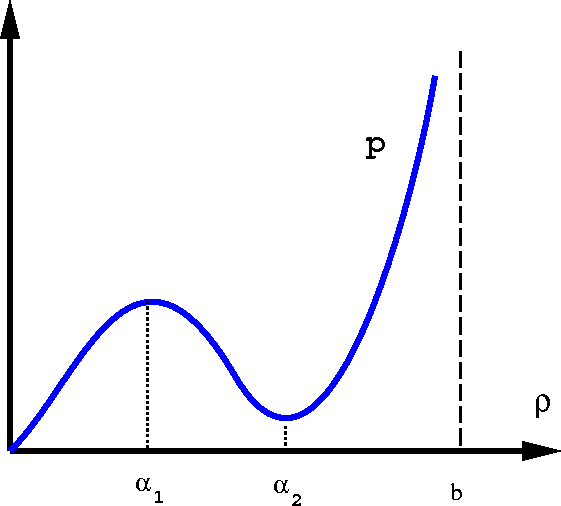}
\label{fig:pressure}
\caption{Pressure $p$ as a function ff density $\rho$ for the values
$b_1=b= 3,\ T_\ast = 0.85$, and $R= 8$.}
\end{figure}

Let an open set $\Omega\subseteq\setR^d,~d\in\{1,2,3\}$ up to the final time $T>0$ be given. The isentropic Navier-Stokes-Korteweg (NSK) equations in arbitrary spatial dimension are given by
\begin{equation}\label{eq:NSK}
\left\{\begin{aligned}
&\partial_t\rho+\nabla\cdot(\rho\vecv)=0,\\
&\partial_t(\rho\vecv)+\nabla\cdot(\rho \vecv\otimes\vecv)+\nabla(p(\rho))=\nabla\cdot{\bf T^\eps[\vecv]}+\gamma\eps^2\rho\nabla\Delta\rho,
\end{aligned}\right.(\vecx,t)\in\Omega\times(0,T),
\end{equation}
where $\rho=\rho(\vecx,t)$ is the density of the fluid, $\vecv=(v_1(\vecx,t),...,v_d(\vecx,t))^T\in\setR^d$ is its momentum and $p(\vecx,t)$ is the pressure.
Note that $\eps$ is the Reynolds number and $\gamma\eps^2$ is the capillary number. We refer to \cite{Jacobs-95,Truskinovsky-94} for a detailed explanation on the physical meaning of the scaling $\eps\to0$ and $\gamma=\mathcal{O}(1)$. 
The matrix $\T^\eps[\vecv]\in\setR^{d\times d}$ in \rf{eq:NSK} 
stands for  the viscous part of the stress tensor which is given 
 for the viscosity coefficients $\nu,\mu\in\setR$ with $\mu\ge 0$ and $3\nu +2\mu >0$
by 
\begin{equation}\label{nsk:stress}
\ds\T^\eps_{ij} :=\ds  \eps \nu \nabla\cdot(\vecv)\delta_{ij} +  2\eps\mu
 \matD_{ij},\quad\matD_{ij}:= \ds\frac{1}{2} \Big(v_{j,x_i}+v_{i,x_j}\Big),\quad (i,j\in\{1,2\}).
\end{equation}
We augment \rf{eq:NSK} with the initial data
\begin{equation}\label{eq:ic_NSK}
\rho(\vecx,0)=\rho_0,\quad \vecv(\vecx,0)=\vecv_0,\quad\vecx\in\Omega,
\end{equation}
and boundary conditions that correspond to a bounded box:
\begin{equation}\label{eq:bc_NSK}
\vecv=0,\quad\nabla\rho\cdot\vecn=0,\quad \vecx\in\partial\Omega.
\end{equation}
To describe a two-phase fluid we choose the Van-der-Waals type pressure
\begin{equation}\label{eq:vdw}
p(\rho)=\frac{RT_\ast\rho}{b-\rho} - b_1\rho^2.
\end{equation} 
Thereby, $b,b_1,R$ are positive constants and $T_\ast$ is the fixed temperature.
If $T_\ast$ is chosen small enough, the pressure $p$ is  monotone decreasing in some non-empty density interval.
This structure allows one to  define phases.
If the density $\rho$ lies in the interval  $(0,\alpha_1]$, ($(\alpha_1,\alpha_2)$), \{$ [\alpha_2,b)$\}
the corresponding fluid state is called {\it vapor (spinodal) \{liquid\}}, see Figure \ref{fig:pressure} for an illustration. 

We observe that the first-order part of \rf{eq:NSK} is not purely hyperbolic for all density values. Indeed, consider, for instance, the one-dimensional (1-D) case, $d=1$. It is easy to check that the eigenvalues of the Jacobian of the first order part of \rf{eq:NSK} $D\vecf^1(\rho,\rho v) \in \setR^{2\times 2}$  are 
\begin{equation}
\lambda_{1}(\rho,u)=v- \sqrt{ p'(\rho)},\quad
\lambda_{2}(\rho,u)=v+ \sqrt{ p'(\rho)}
\end{equation}
with corresponding eigenvectors 
\begin{equation}\label{r:eigenvectorseuler}
  {\bf K}_{1}(\rho,v)=\begin{pmatrix} 1\\ v-\sqrt{p'(\rho)}\end{pmatrix},\quad
  {\bf K}_{2}(\rho,v)=\begin{pmatrix} 1\\ v+\sqrt{p'(\rho)}\end{pmatrix}.
\end{equation}
Therefore, the the first order part of \rf{eq:NSK}  is hyperbolic if and  only if  $\rho\in (0,b) \setminus [\alpha_1,\alpha_2]$.

The lack of hyperbolicity in the first-order part of \rf{eq:NSK} and the presence of the third-order derivative in the momentum balance make the numerical solution of \rf{eq:NSK} to be a challenging task: Explicit schemes suffer from extremely small time steps  while implicit discretizations lead to badly conditioned algebraic problems.  The non-monotonicity of $p$ prevents the use of most modern shock-capturing schemes.  
\subsection{A relaxation for the Navier-Stokes-Korteweg system}
To overcome some of the shortcomings of the classical NSK system we propose a relaxation for the NSK system \cite{Rohde-10}
\begin{equation}\label{eq:aNSK}
\hspace*{-0.2cm}\left\{\begin{aligned}
&\partial_t\rho^{\alpha}+\nabla\cdot(\rho^{\alpha}\vecv^{\alpha})=0,\\
&\partial_t(\rho^{\alpha}\vecv^{\alpha})+\nabla\cdot(\rho^{\alpha} \vecv^{\alpha}\otimes\vecv^{\alpha})+\nabla p(\rho^{\alpha})=\nabla\cdot{\bf T^\eps[\vecv^{\alpha}]}+\frac{1}{\alpha^2}\rho^{\alpha}\nabla(c^{\alpha}-\rho^{\alpha}),\\
&{\gamma\eps^2}\Delta c^{\alpha}+\frac{1}{\alpha^2}(\rho^{\alpha}-c^{\alpha})=0.
\end{aligned}\right.
\end{equation}
Here $\alpha>0$ is the \textit{Korteweg parameter} and $c^{\alpha}$ is a new unknown, that is defined by the additional Poisson equation, and the stress tensor $\T^\eps$ given by \rf{nsk:stress}. The relaxation system \rf{eq:aNSK} is augmented with the initial conditions
\begin{equation}\label{eq:ic_aNSK}
\rho^{\alpha}(\vecx,0)=\rho_0,\quad
\vecv^{\alpha}(\vecx,0)=\vecv_0,\quad\vecx\in\Omega,
\end{equation}
and the boundary conditions
\begin{equation}\label{eq:bc_aNSK}
\vecv^{\alpha}=0,\quad
\nabla c^{\alpha}\cdot\vecn=0,\quad \vecx\in\partial\Omega.
\end{equation}
The system \rf{eq:aNSK} has a structural advantage that becomes evident when we rewrite the time-dependent equations as 
\begin{equation}\label{eq:aNSK_mod}
\left\{\begin{aligned}
&\partial_t\rho^{\alpha}+\nabla\cdot(\rho^{\alpha}\vecv^{\alpha})=0,\\
&\partial_t(\rho^{\alpha}\vecv^{\alpha})+\nabla\cdot(\rho^{\alpha} \vecv^{\alpha}\otimes\vecv^{\alpha})+\nabla \left(p(\rho^\alpha)+\frac{1}{2\alpha^2}\rho^2\right)=\nabla\cdot{\bf T}^\eps[\vecv^{\alpha}]+\frac{1}{\alpha^2}\rho^{\alpha}\nabla c^{\alpha}.
\end{aligned}\right.
\end{equation}
Again, if we consider the 1-D case for the sake of simplicity, then the eigenvalues of the Jacobian of the first-order part of \rf{eq:aNSK_mod}
$ D\vecf^1(\rho^\alpha,\rho^\alpha v^\alpha) \in \setR^{2\times 2}$ are 
\begin{equation}\label{eq:eigenvsluesansk}
\begin{aligned}
&\lambda_{1}(\rho^\alpha,\rho^\alpha v^\alpha)=v^\alpha- \sqrt{ p'(\rho^\alpha)+\frac{1}{\alpha^2}\rho^\alpha},\\
&\lambda_{2}(\rho^\alpha,\rho^\alpha v^\alpha)=v^\alpha+ \sqrt{ p'(\rho^\alpha)+\frac{1}{\alpha^2}\rho^\alpha},
\end{aligned}
\end{equation}
with corresponding eigenvectors 
\begin{equation}\label{eq:eigenvectorsansk}
\begin{aligned}
&{\bf K}_{1}(\rho^\alpha,\rho^\alpha v^\alpha)=\begin{pmatrix} 1\\ v^\alpha-\sqrt{p'(\rho^\alpha)+\frac{1}{\alpha^2}\rho^\alpha}\end{pmatrix},\\
&  {\bf K}_{2}(\rho^\alpha,\rho^\alpha v^\alpha)=\begin{pmatrix} 1\\ v^\alpha+\sqrt{p'(\rho^\alpha)+\frac{1}{\alpha^2}\rho^\alpha}\end{pmatrix}.
\end{aligned}
\end{equation}
A straightforward computation (see \cite{Solci-02}) shows that we obtain a purely hyperbolic system for
\begin{equation}
\frac{1}{\alpha^2}>|{\rm min}\{p'(s)~:~s\in(\alpha_1,\alpha_2)|,
\end{equation}
where $(\alpha_1,\alpha_2)$ is the interval of the decreasing pressures, i.e., $p'(s)<0$, see Figure \ref{fig:pressure}. The system \rf{eq:aNSK}--\rf{eq:bc_aNSK} can be seen as an approximation of the classical NSK system
 with  $(\rho^{\alpha},\rho^\alpha\vecv^\alpha,c^{\alpha})\to (\rho,\rho\vecv,\rho)$ for the {\emph{Korteweg limit}}  $\alpha\to 0$, where  $(\rho,\rho\vecv)$
is the solution of the corresponding initial boundary value problem \rf{eq:NSK}, \rf{eq:ic_NSK}, \rf{eq:bc_NSK}. 
We refer to \cite{Charve13,Charve-13,Corli-Rohde-12,Giesselmann14a,GLT15} for first rigorous results on the Korteweg limit.
It is possible to show that \rf{eq:aNSK} formally converges to \rf{eq:NSK}.
We take the asymptotic expansion
\begin{equation}
\begin{array}{l}
\rho^\alpha=\rho\uz+\alpha\rho\uo+\alpha^2\rho\ut+\dotsc,\\[1.5ex]
\rho^\alpha\vecv^\alpha=\rho\uz\vecv\uz+\alpha\rho\uo\vecv\uo+\alpha^2\rho\ut\vecv\ut+\dotsc,\\[1.5ex]
c^\alpha=c\uz+\alpha c\uo+\alpha^2c\ut+\dotsc,
\end{array}
\end{equation}
for small $\alpha$ and look at the balances within the equations of system \rf{eq:aNSK}. 
Therefore we compute the Taylor expansion at $\rho\uz$ for the pressure
\begin{equation}
\label{Taylor}
\begin{aligned}
p(\rho^{\alpha})&\hspace*{-.2cm}=p(\rho\uz)+p'(\rho\uz)(\rho^{\alpha}-\rho\uz)+p''(\rho\uz)(\rho^{\alpha}-\rho\uz)^2+\dots\\
&=p(\rho\uz)+\alpha p'(\rho\uz)(\rho\uo+\alpha\rho\ut+\dots)\\
&\quad+\alpha^2p''(\rho\uz)(\rho\uo+\alpha\rho\ut+\dots)^2+\dots
\end{aligned}
\end{equation}
A short computation leads to the following terms for the different powers of $\alpha$:

$\mathcal{O}(\alpha^{-2}):$
\begin{equation}
\rho\uz=c\uz
\label{eq:epm2}
\end{equation}

$\mathcal{O}(\alpha^{-1}):$
\begin{equation}
\rho\uo=c\uo
\label{eq:epm1}
\end{equation}

$\mathcal{O}(1):$
\begin{equation}
\begin{aligned}
&\rho\uz_t+\nabla\cdot\left(\rho\uz \vecv\uz\right)=0,\\
&\left(\rho\uz \vecv\uz\right)_t+\nabla\cdot\left(\rho\uz\vecv\uz\otimes\vecv\uz\right)+\nabla \left(p\uz\right)=\nabla\cdot{\bf T^\eps[\vecv\uz]}\\
&\qquad+\rho\uz \left(c\ut-\rho\ut\right)+\rho\uo \nabla\left(c\uo-\rho\uo\right)+\rho\ut \nabla\left(c\uz-\rho\uz\right),\\
&\gamma\eps^2 \Delta c\uz=\left(c\ut-\rho\ut\right).
\end{aligned}
\label{eq:ep0}
\end{equation}
We substitute \rf{eq:epm2},\rf{eq:epm1} into \rf{eq:ep0} and obtain
\begin{equation}
\hspace*{-0.5cm}\left\{\begin{aligned}
&\rho\uz_t+\nabla\cdot(\rho\uz \vecv\uz)=0,\\
&(\rho\uz \vecv\uz)_t+\nabla\cdot(\rho\uz\vecv\uz\otimes\vecv\uz)+\nabla p\uz=\eps \Delta \vecv\uz+\gamma\eps^2\rho\uz \nabla\cdot\Delta\rho\uz,
\end{aligned}\right.
\end{equation}
which is the classical Navier-Stokes-Korteweg system, see \rf{eq:NSK}.

We illustrate this result by the following numerical experiment that is taken from \cite{NeusserRohdeSchleper-15}.  

\begin{exa}[Numerical verification of the Korteweg limit]\label{example1}
\end{exa}
We consider the 1-D system \rf{eq:aNSK} with $\eps=0.01$ on interval  $\Omega=(-1,2)$ subject to the boundary conditions \rf{eq:bc_aNSK} and the following initial data
\begin{equation}
\begin{aligned}
\rho_0(x)&=\left\{\begin{array}{ll}0.3, &x\in(0.3,0.6)\cup(0.85,1.05)\\
1.8, &\mbox{otherwise}\end{array}\right.,\\
v_0(x)&=0.
\end{aligned}
\label{eq:initialbubble1D}
\end{equation}
From  the physical point of view, these initial conditions describe two vapor bubbles surrounded by  liquid fluid.
A numerical solution of this initial-boundary value problem, denoted by  
$\vecu^{\alpha}_h=(\rho^{\alpha}_h,   \rho^{\alpha}_h v^{\alpha}_h,c^{\alpha}_h)^T$ and a 
numerical solution of the corresponding classical NSK system, denoted by  
$\vecu_h=(\rho_h,\rho_h v_h)^T$. Both solutions were computed in \cite{NeusserRohdeSchleper-15} using an explicit local discontinuous Galerkin (LDG) method, see, e.g.,  \cite{BassiRebay,CockburnShu-III,Diehl}.
\begin{table}[h]
 \begin{center}
    \begin{tabular}{c||c|c|c|c|c}
    $\quad i$ & 1&2&3&4&5\\\hline \hline
     $ \Delta x=0.005$,~$\alpha^{-2}_i$ =&1&10&100&1000&10000\\\hline \hline
     CPU-time [s]&563&654&787&828&2178\\\hline
     $D^i_h= \|\vecu_{h} -\vecu^\alpha_h\|_{L^2}$&0.25&0.033&2.9e-3&2.4e-4&3.7e-5\\\hline
     EOC$_i=\frac{{\rm ln}(D_h^i/D_h^{i+1})}{{\rm ln}(\alpha_{i+1}/\alpha_i)}$&-&0.879&1.056&1.082&0.812
    \end{tabular}
 \end{center}
  \caption{Discrete $L^2(\Omega)$-distance. The distance decreases as $\alpha$ does. The fifth line contains the  experimental order of convergence (EOC) with respect to $\alpha$ and the third line contains the CPU time.}
    \label{tab:EOC}
\end{table}

The numerical results, presented in Table \ref{tab:EOC}, indicate that the relaxed model is an $\mathcal{O}(\alpha^2)$-approximation of the original system. As one can also see from this Table,  for decreasing values of $\alpha$ the CPU-time is increasing due to the dependence on $\alpha$ of the eigenvalues \rf{eq:eigenvsluesansk}. The maximum wave speed for system \rf{eq:aNSK} is $\lambda_\text{max}=|v^\alpha_{max}|+\sqrt{p'(\rho^\alpha_{max})}$ (where $\rho^\alpha_\text{max}$ and $v^\alpha_\text{max}$ are the maximum values of the density and velocity, respectively).
For an explicit scheme one needs 
\begin{equation}\label{eq:cfl}
\Delta t=c_{DG}\min\left\{\frac{\dx}{\lambda_{max}},\frac{\dx^2}{\eps}\right\}=c_{DG}\min\left\{\frac{\dx}{|v^\alpha_{max}|+\sqrt{p'(\rho^\alpha_{max})}},\frac{\dx^2}{\eps}\right\},
\end{equation}
for some $0\leq c_{DG}<1$ to satisfy the CFL condition for stability. For small $\alpha$ one needs $\dt=\mathcal{O}(\alpha\dx)$. This restriction is a huge drawback for numerical simulations and we are interested in finding a way to circumvent this restriction.


\section{An Asymptotic-Preserving Scheme in the Korteweg Limit}\label{sec:allspeed}

Having in mind the shortcomings of the relaxation system \rf{eq:aNSK}, we propose a new AP numerical scheme. 
Note that from here on we suppress the index $\alpha$ of all of the primal variables in order to shorten the notation.
 
\subsection{A hyperbolic splitting}
The numerical solution of the relaxation system \rf{eq:aNSK} requires a resolution of two scales: the (fast) wave scale, coming from the Korteweg part and the (slow) convection scale. In order to obtain an accurate and efficient numerical scheme, which is able to handle both scales, we implement a splitting approach. To this end we first introduce a parameter $a$ and rewrite the system \rf{eq:aNSK} in the following form (by adding and subtracting $a\rho\nabla\rho$):
\begin{equation}\label{eq:NSK_a}
\left\{\begin{aligned}
&\partial_t\rho+\nabla\cdot(\rho\vecv)=0,\\
&\partial_t(\rho\vecv^{\alpha})+\nabla\cdot(\rho \vecv^\alpha\otimes\vecv)+\nabla\tilde{p}(\rho)=\nabla\cdot{\bf T^\eps[\vecv]}+\frac{1}{\alpha^2}\rho\nabla(c-\rho)+a\rho\nabla\rho,\\
&\gamma\eps^2\Delta c+\frac{1}{\alpha^2}(\rho-c)=0.
\end{aligned}\right.
\end{equation}  
Here the pressure is defined as 
\begin{equation}
\tilde{p}(\rho)=p(\rho)+\frac{a}{2}\rho^2
\end{equation}
with 
\begin{equation}\label{eq:p_a}
a=|{\rm min}\{p'(s)~:~s\in\{\alpha_1,\alpha_2\}\}|,
\end{equation}
where $(\alpha_1,\alpha_2)$ is the interval of the decreasing pressures, i.e., $p'(s)<0$, see Figure \ref{fig:pressure}. 

To resolve the different wave length scales, we then split \eqref{eq:NSK_a} into two systems: The slow dynamics are described by the system 
\begin{equation}\label{eq:slow}
\left\{\begin{aligned}
&\partial_t\rho+\beta\nabla\cdot(\rho\vecv)=0,\\
&\partial_t(\rho\vecv)+\nabla\cdot(\rho \vecv\otimes\vecv)+\nabla\tilde{p}(\rho)=\nabla\cdot{\bf T^\eps[\vecv]},
\end{aligned}\right.
\end{equation}
and for the fast dynamics we have
\begin{equation}\label{eq:fast}
\left\{\begin{aligned}
&\partial_t\rho+(1-\beta)\nabla\cdot(\rho\vecv)=0,\\
&\partial_t(\rho\vecv)=\frac{1}{\alpha^2}\rho\nabla(c-\rho)+a\rho\nabla\rho,\\
&\gamma\eps^2\Delta c +\frac{1}{\alpha^2}(\rho-c)=0.
\end{aligned}\right.
\end{equation}
The splitting parameter $0<\beta<1$ in \rf{eq:slow} and \rf{eq:fast} determines how much of the momentum is seen by each system and the choice of $a$ in \rf{eq:p_a} is motivated by hyperbolicity. In the 1-D case, for instance, the wave speeds are
\[\lambda=v\pm\sqrt{(1-\beta)v^2+\beta\tilde{p}'(\rho)}\]
and when we choose $a$ as in \rf{eq:p_a}, we ensure that the slow system \rf{eq:slow} is hyperbolic and that the wave speeds do not depend on $\alpha$ any more.

In what follows, we present numerical methods used to solve each one of the subsystems. We can discretize the slow system \rf{eq:p_a} using any explicit shock-capturing scheme for the compressible Navier-Stokes equations noting that the wave speeds are no longer stiff which avoids time step problems seen in the original system \rf{eq:NSK_a}.
A stability analysis shows that we need to fulfil the following CFL condition for the slow system:
\begin{equation}\label{eq:cflap}
\begin{array}{lll}
\Delta t&=&\ds c_{AP}\min\left\{\frac{\dx}{\lambda_{max}},\frac{\dx^2}{\eps}\right\}\\[1.5ex]
&=&\ds c_{AP}\min\left\{\frac{\dx}{|v_{max}|+\sqrt{(1-\beta)v_{max}^2+\beta\tilde{p}'(\rho_{max})}},\frac{\dx^2}{\eps}\right\}.
\end{array}
\end{equation}
for some $0\leq c_{AP}<1$.
For the fast dynamics governed by \rf{eq:fast}, we introduce an implicit scheme that corresponds to the discretization of a linear hyperbolic system with variable coefficients and therefore does not have any time step restrictions.

\subsection{Time discretizations of the split schemes}
In this section, we outline the first-order in time discretizations for the systems \rf{eq:slow} and \rf{eq:fast}. Note that the MUSCL methodology allows us to turn any first order scheme into a second order one. 
As mentioned in the previous section, we choose an explicit time discretization for the slow dynamics
\begin{equation}\label{eq:slowt}
\begin{aligned}
&{\displaystyle\frac{\rp-\rn}{\dt}}+\beta\nabla\cdot(\rn\un)=0,\\
&{\displaystyle\frac{\rup-\run}{\dt}}+\nabla\cdot(\run\otimes\un)+\nabla\tilde{p}(\rn)=\nabla\cdot{\bf T^\eps[\un]},
\end{aligned}
\end{equation}
and the following implicit-explicit discretization for the fast dynamics:
\begin{equation}\label{eq:fastt}
\begin{aligned}
&{\displaystyle\frac{\rp-\rn}{\dt}}+(1-\beta)\nabla\cdot(\rup)=0,\\
&{\displaystyle\frac{\rup-\run}{\dt}}=\frac{1}{\alpha^2}\rn\nabla(\cp-\rp)+a\rn\nabla\rp,\\
&\gamma\eps^2\Delta \cp +\frac{1}{\alpha^2}(\rp-\cp)=0.
\end{aligned}
\end{equation}
The time discretization of the nonlinear terms in the fast dynamics system \rf{eq:fastt} is one of the key ingredients of our scheme. With this choice, the system for the fast dynamics is a linear hyperbolic system with constant coefficients at each time step. We show the benefit of this choice further below.

Now we follow the idea of \cite{Haack-12} to obtain an efficient numerical solution strategy. We sum up \rf{eq:slowt} and \rf{eq:fastt} to have
\begin{equation}\label{eq:im1}
\begin{aligned}
&{\displaystyle\frac{\rp-\rn}{\dt}}+\beta\nabla\cdot(\run)+(1-\beta)\nabla\cdot(\rup)=0,\\
&{\displaystyle\frac{\rup-\run}{\dt}}+\nabla\cdot(\rn \un\otimes\un)+\nabla\tilde{p}(\rn)\\
&\hspace{2cm}=\nabla\cdot{\bf T^\eps[\un]}+\frac{1}{\alpha^2}\rn\nabla(\cp-\rp)+a\rn\nabla\rp,\\
&\gamma\eps^2\Delta\cp+\frac{1}{\alpha^2}(\rp-\cp)=0,
\end{aligned}
\end{equation}
and rewrite the third equation in \rf{eq:im1} as
\begin{equation}\label{eq:im2}
\rp=\cp-{\gamma\eps^2}\alpha^2\Delta\cp.
\end{equation}
We then substitute \rf{eq:im2} into the momentum equation of \rf{eq:im1} to get
\begin{equation}\label{eq:im3}
\begin{aligned}
{\displaystyle\frac{\rup-\run}{\dt}}&=-\nabla\cdot(\rn \un\otimes\un)-\nabla\tilde{p}(\rn)-\nabla\cdot{\bf T^\eps[\un]}\\
&+\frac{1}{\alpha^2}\rn\nabla(\cp-\cp+{\gamma\eps^2}\alpha^2\Delta\cp)\\
& +a\rn\nabla(\cp-{\gamma\eps^2}\alpha^2\Delta\cp).
\end{aligned}
\end{equation}
Next, we solve \rf{eq:im3} for $\rup$ and obtain
\begin{equation}\label{eq:im4}
\begin{aligned}
\rup&=\run-\dt\nabla\cdot(\rn \un\otimes\un)-\dt\nabla\tilde{p}(\rn)\\
&+\dt\nabla\cdot{\bf T^\eps[\un]}+\dt\gamma\eps^2\rn\nabla\Delta\cp\\
&+a\dt\rn\nabla(\cp-{\gamma\eps^2}\alpha^2\Delta\cp).
\end{aligned}
\end{equation}
Finally, we substitute \rf{eq:im4} into the density equation of \rf{eq:im1}, obtaining a plate equation for $\cp$:
\begin{equation*}
\begin{aligned}
&{\displaystyle\frac{\cp-\gamma\eps^2\alpha^2\Delta\cp-\rn}{\dt}}+\beta\nabla\cdot(\run)+(1-\beta)\nabla\cdot\run\\
&\hspace*{1cm}+(1-\beta)\dt\nabla\cdot\left[-\nabla\cdot(\rn \un\otimes\un)-\nabla\tilde{p}(\rn)\right.\\
&\hspace*{1cm}+{\bf T^\eps[\un]}+\gamma\eps^2\rn\nabla\Delta\cp \left.+a\left(\rn\nabla\cp-\gamma\eps^2\alpha^2\rn\nabla\Delta\cp\right)\right]=0.
\end{aligned}
\end{equation*}
%
The last equation is now a plate equation with respect to the unknown variable $\cp$ and the terms from the previous steps can be pushed to the right hand side as source terms. The crucial part is the discretization of the non-conservative terms. With the discretization we propose, we have to solve a symmetric, sparse linear system for $\cp$. In two space dimensions, the biharmonic operator is a tridiagonal matrix ${\bf M} \in\setR^{M\times N}$ with bandwidth $\mathcal{O}(N)$, where $N(M)$ is the number of grid cells in the $x$-($y$-)direction. This system can for example be  solved with a conjugate gradient method. Note that we do not compute the inverse matrix $M^{-1}$, but solve the linear system directly. As long as $N,M$ are not too large, this strategy showed to be very efficient in our numerical test cases. The updated density $\rp$ and momentum $\rup$ are obtained from \rf{eq:im2}, \rf{eq:im4}. 

Before we depict the spatial discretization, we want to sum up the key ideas of our numerical scheme.
We split up the relaxation system into two smaller system to resolve the two wave scales. With our choice of the splitting parameter $a$, the time steps of the explicit scheme for the slow system are independent of $\alpha$.
Additionally, we proposed a linearization of the fast system, that leads to an implicit scheme. 
After some small computations we have also showed, that evolution in time corresponds to the solution of a plate equation. 

\subsection{Spatial discretization of the split systems}
We assume a two-dimensional (2-D), rectangular grid with uniform spacing $\Delta x$ for ease of explanation. Let $\mathcal{T}=\{B_{i,j}|i=1,\dots ,N;~j=1,\dots ,M\}$ be a partition of $\Omega$  where $B_{i,j}$ is a square with length $\dx$ and $N (M)$ is the number of cells in the $x$-($y$-)direction. We define $\phi_{i,j}=\phi(x_i,y_j)$ where $(x_i,y_j)=(-\dx/2+i\dx,-\dy/2+j\dy)$. Within this section we define $\vecu=(\rho,\rho \vecv)^T$. 

\subsection{Discretization of the slow system}
We rewrite system \rf{eq:slowt} as
\begin{equation}
\frac{\vecu^{n+1}-\vecu^n}{\dt}+\vecf^1(\vecu)_x+\vecf^2(\vecu)_y=\nabla\cdot{\bf T^\eps[\vecv^n]}
\label{system1}
\end{equation}
with $\vecu^n=(\rho^n,\rho^n u^n,\rho^n v^n)^T$ and $\vecf^{1}(\vecu)$, $\vecf^{2}(\vecu)$ defined as
\[
\vecf^1(\vecu)=(\beta\rho u,\rho u^2+ \tilde{p}(\rho),\rho u v)^T,\quad \vecf^2(\vecu)=(\beta\rho v,\rho u v,\rho v^2+ \tilde{p}(\rho))^T.
\]
First, we treat the convective flux terms and thus recall that the eigenvalues of the Jacobians $D\vecf^1,~D\vecf^2$ are given by
\begin{equation}
\begin{aligned}
&\lambda^1_1(\vecu)=u-\sqrt{(1-\beta)u^2+\beta \tilde{p}'(\rho)},\\
&\lambda^1_2(\vecu)=u,\\
&\lambda^1_3(\vecu)=u+\sqrt{(1-\beta)u^2+\beta \tilde{p}'(\rho)},\\[2ex]
&\lambda^2_1(\vecu)=v-\sqrt{(1-\beta)v^2+\beta \tilde{p}'(\rho)},\\
&\lambda^2_2(\vecu)=v,\\
&\lambda^2_3(\vecu)=v+\sqrt{(1-\beta)v^2+\beta \tilde{p}'(\rho)}.
\end{aligned}
\end{equation}
We discretize the fluxes $\vecf^1$ and $\vecf^2$ terms using a HLL-solver \cite{Torobook}, so that the numerical fluxes are given by
\begin{equation}
\begin{aligned}
&\matH^1(\vecu^-,\vecu^+) = \ds\frac{\lambda^1_{+}\vecf^1(\vecu^-) - \lambda^1_{-}\vecf^1(\vecu^+)
							  - \lambda^1_{+}\lambda^1_{-}(\vecu^+-\vecu^{-})}
							  {\lambda^1_{+}-\lambda^1_{-}},\\
&\matH^2(\vecu^-,\vecu^+) = \ds\frac{\lambda^2_{+}\vecf^2(\vecu^-) - \lambda^2_{-}\vecf^2(\vecu^+)
							  - \lambda^2_{+}\lambda^2_{-}(\vecu^+-\vecu^{-})}
							  {\lambda^2_{+}-\lambda^2_{-}}.
\end{aligned}
\label{flux}
\end{equation}
with $\lambda^i_\pm$ defined as
\begin{equation}
\begin{array}{l}
\lambda^i_+:=\lambda^i_+(\vecu^-,\vecu^+)=\max\{\lambda^i_3(\vecu^-),\lambda^i_3(\vecu^+),0\},\\[1.5ex]
\lambda^i_-:=\lambda^i_-(\vecu^-,\vecu^+)=\min\{\lambda^i_1(\vecu^-),\lambda^i_1(\vecu^+),0\}.
\end{array}
\end{equation}
for $i\in\{1,2\}$.

With the numerical fluxes \rf{flux} we define the discrete conservative operator for the convective part $\vecf^1(\vecu)_x+\vecf^2(\vecu)_y$ of \rf{system1}
\begin{equation}\label{eq:conservative}
	\begin{array}{l}
\displaystyle\mathcal{F}(\vecu\ijn)=\frac{\matH_{i+\frac{1}{2},j}^{n}-\matH_{i-\frac{1}{2},j}^{n}}{\dx}+
\frac{\matH_{i,j+\frac{1}{2}}^{n}-\matH_{i,j-\frac{1}{2}}^{n}}{\dy},
\end{array}
\end{equation}
where 
\begin{equation}
\matH_{i+\frac{1}{2},j}^{n} = \matH^1(\vecu_{i+\frac{1}{2},j}^{n,-},\vecu_{i+\frac{1}{2},j}^{n,+}),\qquad
\matH_{i,j+\frac{1}{2}}^{n} = \matH^2(\vecu_{i,j+\frac{1}{2}}^{n,-},\vecu_{i,j+\frac{1}{2}}^{n,+}).
\end{equation}
The values of $\vecu_{i\pm1/2,j}$ at the cell interfaces are reconstructed component-wise using the generalized minmod limiter with $\theta\in[1,2]$,
\begin{equation}
\begin{aligned}
&\vecu_{i+\frac{1}{2},j}^{n,+}=\vecu\pjn-\frac{\dx}{2}{\bf\sigma}\pjn,\qquad
\vecu_{i+\frac{1}{2},j}^{n,-}=\vecu\ijn+\frac{\dx}{2}{\bf\sigma}\ijn,\\[2ex]
&{\bf\sigma}\ijn=\operatorname{minmod}\displaystyle\left(\theta\frac{\vecu\pjn-u\ijn}{\dx},\theta\frac{\vecu\ijn-\vecu\mjn}{\dx},\frac{\vecu\pjn-u\mjn}{2\dx}\right).
\end{aligned}
\end{equation}
The values of $\vecu_{i,j\pm1/2}$ at the cell interfaces are reconstructed in a similar manner. The
 proposed scheme \rf{eq:conservative} is general and may be used in conjunction with one's favorite numerical flux replacing the HLL-flux \rf{flux}.

The viscous part of the stress tensor can be written as
\begin{equation}
\nabla\cdot{\bf T^\eps[\vecv]}=\eps\begin{pmatrix}
\xi(u_x+v_y)_x+\mu(u_y-v_x)_y\\
-\mu(u_y-v_x)_x+\xi(u_x+v_y)_y
\end{pmatrix}.
\end{equation}
Here we set $\xi=2\mu+\nu$ where $\mu,~\nu$ denote the coefficients of viscosity. We define the discrete central difference operators
\begin{equation}
D_x\phi_{i,j}^n=\frac{\phi\pjn-\phi\mjn}{2\dx},\qquad
 D_y\phi_{i,j}^n=\frac{\phi\ipn-\phi\imn}{2\dy}
\label{central}
\end{equation}
and write the discrete version of the stress tensor
\begin{equation}\label{eq:stress}
\begin{aligned}
\tilde{\nabla}&\cdot{\bf T^\eps[\vecv\ijn]}=\\
&\eps~\begin{pmatrix}
\xi D_x(D_x u\ijp+D_y v\ijp)_{i,j}+\mu D_y(D_y u\ijp-D_x v\ijp)_{i,j}\\
-\mu D_y(D_y u\ijp-D_x v\ijp)_{i,j}+\xi D_x(D_x u\ijp+D_y v\ijp)_{i,j}
\end{pmatrix}
\end{aligned}
\end{equation}
We combine \eqref{eq:conservative},\eqref{eq:stress} together and obtain the following discretization of the system
\rf{system1}:
\begin{equation}\label{eq:slowd}
\begin{aligned}
&\ds\sum_{i,j}\left[\frac{\rho\ijp-\rho\ijn}{\dt}+\mathcal{F}^1(\vecu\ijn)\right]=0\\
&\ds\sum_{i,j}\left[\frac{\rho\ijp\vecv\ijp-\rho\ijn\vecv\ijn}{\dt}+\displaystyle\mathcal{F}^2(\vecu\ijn)-\tilde{\nabla}\cdot{\bf T^\eps[\vecv\ijn]}\right]=0.
\end{aligned}
\end{equation}.

\subsection{Discretization of the fast system}
The discretization of the fast system \eqref{eq:fastt} consists of three parts. First, we discretize the momentum term by using central differences
\begin{equation}\label{eq:central}
\tilde{\nabla}\cdot(\rho\vecv)\ijp=D_x(\rho\ijp u\ijp)+D_y(\rho\ijp v\ijp),
\end{equation}
with the operators $D_x$ and $D_y$ defined in \rf{central}.
Then we discretize the elliptic term as
\begin{equation}\label{eq:elliptic}
\tilde{\Delta} c\ijp=\frac{D_x c\pjp-D_x c\mjp}{2\dx}+\frac{D_y c\ipp+D_y c\imp}{2\dy}.
\end{equation}
The terms $\rho^n\Delta\phi$ are non-conservative terms. The discretization of these terms is by no means unique and we chose for our scheme
\begin{equation}\label{eq:nc}
(\rho\ijn\tilde{\nabla} \phi\ijp)_{nc}=\frac{\rho\ijn}{2}\begin{pmatrix}
\frac{\phi\pjp-\phi\mjp}{\dx}\\
\frac{\phi\ipp-\phi\imp}{\dy}
\end{pmatrix}.
\end{equation}
We combine \eqref{eq:central}, \eqref{eq:elliptic}, and \eqref{eq:nc} together and obtain the following discretization for the fast dynamics  \eqref{eq:fastt}:
\begin{equation}\label{eq:fastd}
\begin{aligned}
&\ds\sum_{i,j}\left[\frac{\rho\ijp-\rho\ijn}{\dt}+(1-\beta)\tilde{\nabla}\cdot(\rho\vecv\ijp)\right]=0,\\
&\ds\sum_{i,j}\left[\frac{\rho\ijp\vecv\ijp-\rho\ijn\vecv\ijn}{\dt}-
\frac{1}{\alpha^2}(\rho\ijn\tilde{\nabla}  c\ijp)_{nc}
+(\frac{1}{\alpha^2}-a)(\rho\ijn\tilde{\nabla} \rho\ijp)_{nc}\right]=0,\\
&\ds\sum_{i,j}\left[\gamma\eps^2\tilde{\Delta} c\ijp +\frac{1}{\alpha^2}(\rho\ijp-c\ijp)\right]=0.
\end{aligned}
\end{equation}

\subsection{Discretization of the NSK system}
For the sake of completeness we also provide the discretization of the NSK system \rf{eq:NSK}. We use the same ideas and notations as in the previous chapters. For the slow dynamics the discretization reads:
\begin{equation}\label{eq:slowdnsk}
\begin{aligned}
&\ds\sum_{i,j}\left[\frac{\rho\ijp-\rho\ijn}{\dt}+\mathcal{F}^1(\vecu\ijn)\right]=0,\\
&\ds\sum_{i,j}\left[\frac{\rho\ijp\vecv\ijp-\rho\ijn\vecv\ijn}{\dt}+\displaystyle\mathcal{F}^2(\vecu\ijn)-\tilde{\nabla}\cdot{\bf T^\eps[\vecv\ijn]}\right]=0,
\end{aligned}
\end{equation}
and for the fast dynamics we have
\begin{equation}\label{eq:fastdnsk}
\begin{aligned}
&\ds\sum_{i,j}\left[\frac{\rho\ijp-\rho\ijn}{\dt}+(1-\beta)\tilde{\nabla}\cdot(\rho\vecv\ijp)\right]=0,\\
&\ds\sum_{i,j}\left[\frac{\rho\ijp\vecv\ijp-\rho\ijn\vecv\ijn}{\dt}-\gamma\eps^2(\rho\ijn\tilde{\nabla}\tilde{\Delta} \rho\ijp)_{nc}-a(\rho\ijn\tilde{\nabla} \rho\ijp)_{nc}\right]=0.
\end{aligned}
\end{equation}

\subsection{Boundary conditions}
We have to account for the boundary conditions \rf{eq:bc_NSK}, \rf{eq:bc_aNSK} and enforce this conditions by using ghost cells. We introduce an associated cell at the edge of each element  $B_{i,j}$ which is part of $\partial\Omega$.
For scheme \rf{eq:slowd}, \rf{eq:fastd} we set
\[
\begin{array}{rlrlrlrl}
\rho_{0,j}&=\rho_{1,j},&\quad u_{0,j}&=-u_{1,j},&\quad v_{0,j}&=v_{1,j},&\quad c_{0,j}&=c_{1,j},\\
\rho_{N+1,j}&=\rho_{N,j},&\quad u_{N+1,j}&=-u_{N,j},&\quad v_{N+1,j}&=v_{N,j},&\quad c_{N+1,j}&=c_{N,j},\\
\rho_{i,0}&=\rho_{i,1},&\quad u_{i,0}&=u_{i,1},&\quad v_{i,0}&=-v_{i,1},&\quad c_{i,0}&=c_{i,1},\\
\rho_{i,M+1}&=\rho_{i,M},&\quad u_{i,M+1}&=u_{i,M},&\quad v_{i,M+1}&=-v_{i,M},&\quad c_{i,M+1}&=c_{i,M},
\end{array}
\]
and for scheme \rf{eq:slowdnsk}, \rf{eq:fastdnsk} we set
\[
\begin{array}{rlrlrl}
\rho_{0,j}&=\rho_{1,j},&\quad u_{0,j}&=-u_{1,j},&\quad v_{0,j}&=v_{1,j},\\
\rho_{N+1,j}&=\rho_{N,j},&\quad u_{N+1,j}&=-u_{N,j},&\quad v_{N+1,j}&=v_{N,j},\\
\rho_{i,0}&=\rho_{i,1},&\quad u_{i,0}&=u_{i,1},&\quad v_{i,0}&=-v_{i,1},\\
\rho_{i,M+1}&=\rho_{i,M},&\quad u_{i,M+1}&=u_{i,M},&\quad v_{i,M+1}&=-v_{i,M}.
\end{array}
\]
\subsection{Asymptotic preserving property}
In the previous section we provided numerical schemes for the Navier-Stokes-Korteweg system and for its relaxation. 
This allows use to formulate the main theorem of this work.  
\begin{thm}
The fully discrete numerical scheme \rf{eq:slowd},\rf{eq:fastd} is asymptotic preserving in the Korteweg limit $\alpha\to0$ in the sense that it transforms into a consistent discretization of the Navier-Stokes-Korteweg system \rf{eq:slowdnsk},\rf{eq:fastdnsk} as $\alpha\to0$.
\end{thm}
Proof:

We start with the asymptotic expansion with respect to a small parameter $\alpha$:
\[\begin{array}{l}
\rho\ijn=\rho\nz+\alpha\rho\no+\alpha^2\rho\nt+\dotsc,\\[1.5ex]
\rho\ijp=\rho\pz+\alpha\rho\po+\alpha^2\rho\pt+\dotsc,\\[1.5ex]
\rho\ijn\vecv\ijn=\rho\nz\vecv\nz+\alpha\rho\no\vecv\no+\alpha^2\rho\nt\vecv\nt+\dotsc,\\[1.5ex]
\rho\ijp\vecv\ijp=\rho\pz\vecv\pz+\alpha\rho\po\vecv\po+\alpha^2\rho\pt\vecv\pt+\dotsc,\\[1.5ex]
 \tilde{p}\ijn= \tilde{p}\nz+\alpha \tilde{p}\no+\alpha^2 \tilde{p}\nt+\dotsc,\\[1.5ex]
c\ijp=c\pz+\alpha c\po+\alpha^2 c\pt+\dotsc,
\end{array}
\]
and look at the balances within the equations \rf{eq:slowd},\rf{eq:fastd}.
At  $\mathcal{O}\left(\alpha^{-2}\right)$, we have the balance
\begin{equation}
\rho\pz=c\pz,
\label{eq:depm2}
\end{equation}
and at $\mathcal{O}\left(\alpha^{-1}\right)$ we have
\begin{equation}
\rho\po=c\po.
\label{eq:depm1}
\end{equation}
For the $\mathcal{O}\left(1\right)$ terms we compute for the slow system
\begin{equation}
\begin{array}{l}
\ds\frac{\rho\pz-\rho\nz}{\dt}+\mathcal{F}^1\left(\vecu\nz\right)=0,\\[3ex]
\ds\frac{\rho \vecv\pz-\rho \vecv\nz}{\dt}+\mathcal{F}^2\left(\vecu\nz\right)=\tilde{\nabla}\cdot{\bf T^\eps[\vecv\nz]},
\end{array}
\label{eq:dep1}
\end{equation}
and the fast system
\begin{equation}
\begin{array}{l}
\ds\frac{\rho\pz-\rho\nz}{\dt}+\left(1-\beta\right)\tilde{\nabla}\cdot\left(\rho\vecv\pz\right)=0,\\[3ex]
\ds\frac{\rho u\pz-\rho u\nz}{\dt}=\left(\rho\nz\tilde{\nabla}  \left(c\pt-\rho\pt\right)\right)_{nc}+a\left(\rho\nz\tilde{\nabla}\rho\pz\right)_{nc}\\[2ex]
\hspace{1.2cm}+\left(\rho\no\tilde{\nabla}  \left(c\po-\rho\po\right)\right)_{nc}+\left(\rho\nt\tilde{\nabla}  \left(c\pz-\rho\pz\right)\right)_{nc},\\[4ex]
\gamma\eps^2\tilde{\Delta} c\pz +\left(\rho\pt-c\pt\right)=0.
\end{array}
\label{eq:dep2}
\end{equation}
Now we use \rf{eq:depm2},\rf{eq:depm1} and the third equation in \rf{eq:dep2} and finally obtain for the slow system
\begin{equation}
\setlength\mathindent{0cm}
\begin{array}{l}
\ds\frac{\rho\pz-\rho\nz}{\dt}+\mathcal{F}^1\left(\vecu\nz\right)=0,\\[3ex]
\ds\frac{\rho \vecv\pz-\rho \vecv\nz}{\dt}+\mathcal{F}^2\left(\vecu\nz\right)=\tilde{\nabla}\cdot{\bf T^\eps[\vecv\nz]},
\end{array}
\label{eq:dep3}
\end{equation}
and for the fast system
\begin{equation}
\begin{array}{l}
\ds\frac{\rho\pz-\rho\nz}{\dt}+\left(1-\beta\right)\tilde{\nabla}\cdot\left(\rho\vecv\pz\right)=0,\\[3ex]
\ds\frac{\rho \vecv\pz-\rho \vecv\nz}{\dt}=\gamma\eps^2\left(\rho\nz\tilde{\nabla} \tilde{\Delta}\rho\pz\right)_{nc} +a\left(\rho\nz\tilde{\nabla}\rho\pz\right)_{nc}.
\end{array}
\label{eq:dep4}
\end{equation}
This is the numerical scheme that we derived in \eqref{eq:slowdnsk}, \rf{eq:fastdnsk}. This means that the scheme \eqref{eq:slowd}, \rf{eq:fastd} is an AP scheme for $\alpha\rightarrow0$.

\section{Numerical Experiments for the asymptotic-preserving scheme}\label{sec:numerics}
In this section, we present a series of experiments that compare the performance of the AP scheme presented in the previous section with the results obtained by using the explicit discontinuous Galerkin scheme from \cite{NeusserRohdeSchleper-15}. The time step $\Delta T$ was chosen according to the CFL condition \eqref{eq:cfl} for the discontinuous Galerkin scheme with $c_{DG}$ and according to CFL condition \eqref{eq:cflap} with $c_{AP}=0.4$ for the AP scheme in all numerical computations.


\subsection{Computational efficiency}
\begin{figure}[ht!]
\subfigure[t=0]{\includegraphics[width=0.3\textwidth]{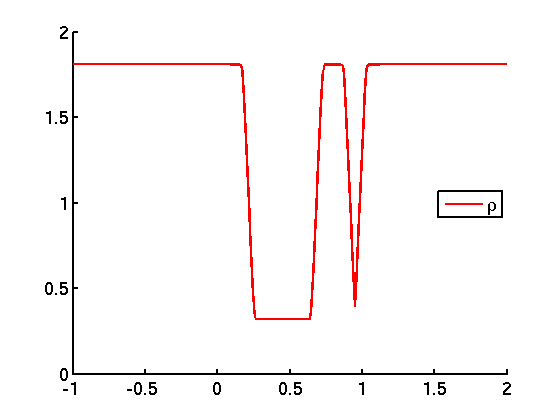}}
\subfigure[t=0.02]{\includegraphics[width=0.3\textwidth]{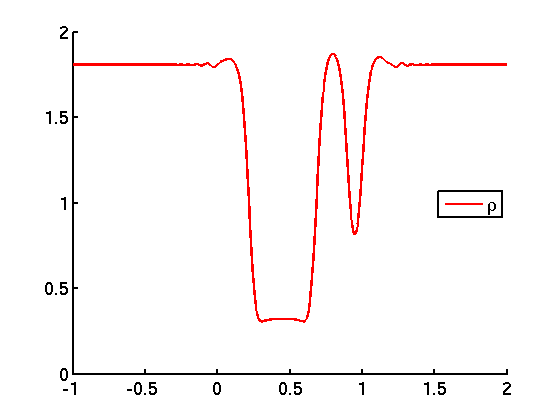}} 
\subfigure[t=0.04]{\includegraphics[width=0.3\textwidth]{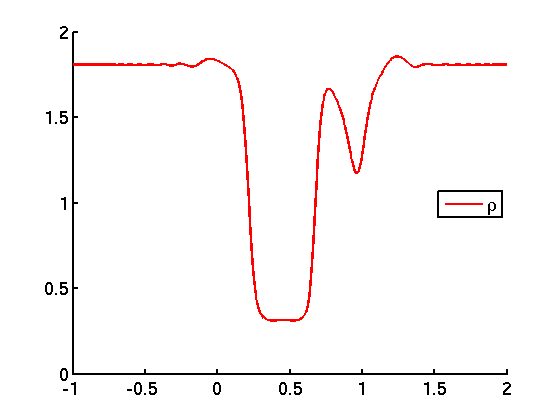}}\\
\subfigure[t=0.4]{\includegraphics[width=0.3\textwidth]{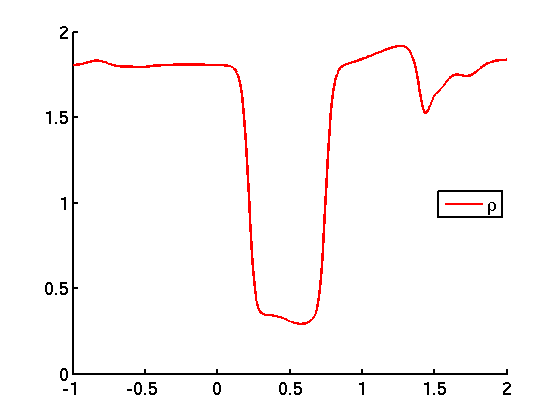}}
\subfigure[t=1]{\includegraphics[width=0.3\textwidth]{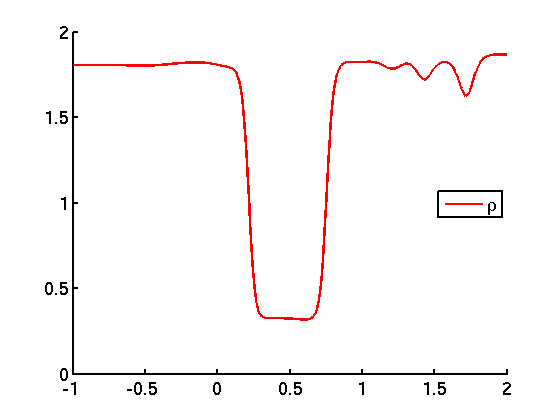}}
\subfigure[t=4]{\includegraphics[width=0.3\textwidth]{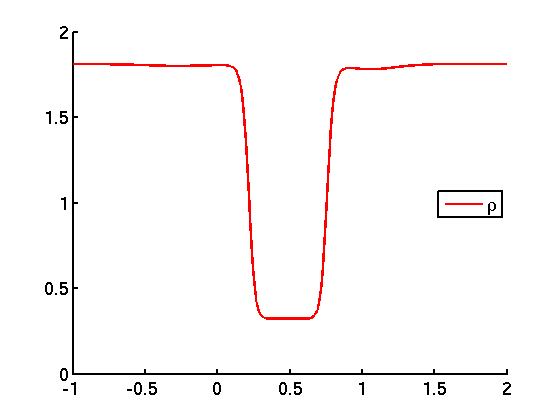}}
\caption{Density  evolution for the relaxation system \rf{eq:NSK_a} and $\alpha=100$  with initial datum \rf{eq:initialbubble1D}. The initial configuration at t=0 corresponds to two bubbles in a box filled with liquid. The smaller bubble shrinks (t=0.02, t=0.04), and emits a shock wave as it collapses (t=0.4, t=1) . At t=4 the material seems to be in equilibrium.}
\label{fig:Bubbles1D}
\end{figure}

In this test case, we consider two different numerical schemes for the relaxation system \eqref{eq:NSK_a}. 
We set $d=1,~\gamma=0.16,~\gamma\eps^2=10^{-5}$ and start with initial conditions 
\begin{equation}
\begin{aligned}
\rho_0(x)&=\left\{\begin{array}{ll}0.3, &x\in(0.3,0.6)\cup(0.85,1.05)\\
1.8, &\mbox{otherwise}\end{array}\right.,\\
v_0(x)&=0.
\end{aligned}
\label{eq:initialbubble1D1}
\end{equation}
that correspond to a two-phase density distribution. Figure \ref{fig:Bubbles1D} shows the evolution of the density at different times that was computed with the AP scheme \eqref{eq:slowd}, \eqref{eq:fastd}.

We compare the performance of the AP scheme and Discontinuous Galerkin scheme of polynomial order 1, which was introduced in \cite{NeusserRohdeSchleper-15}. We use the two schemes to approximate solutions of the NSK system \eqref{eq:NSK}. 

Tables \ref{tab:IMEXerror} and \ref{tab:DGerror} provide numerical evidence to the properties of the our scheme that we stated in Section \ref{sec:allspeed}. We take a column wise look at the CPU-times in Table \ref{tab:IMEXerror} and notice that they are constant for all values of $\alpha$ and for fixed $\Delta x$. This means that for this test case the CPU time is independent of the parameter $\alpha$. In contrast, the CPU-time increases as $\alpha$ decreases for fixed $\Delta x$ in Table \ref{tab:DGerror}. For example the  AP scheme is faster by a factor 73  for $\alpha=10000$ and $\dx=0.005$ and even by a factor 84 for $\alpha=100000$ and $\dx=0.00125$. 

Now we look at the discrete $L_2$-distances for different $\alpha$ and $\Delta x$ compared to a reference solution $\bar{\vecu}$, that was computed with an fully implicit DG-scheme of polynomial order 1 for system \rf{eq:NSK} at $\dx=0.000625$. The last line in Table \ref{tab:DGerror} displays the discrete $L_2$-distance for a fully implicit DG-scheme of polynomial order 1 for system \rf{eq:NSK}. Tables \ref{tab:IMEXerror} and \ref{tab:DGerror} allow to make three statements on the $L_2$-errors for this test case.
First, the $L_2$-error for fixed $\alpha$ decreases as $\dx$ does for both schemes.
Secondly, the discrete $L_2$-distances decreases for fixed $\dx$ as $\alpha$ does for both schemes.
Thirdly, the $L_2$-error for each $\alpha$ and $\dx$ for the DG scheme is slightly better than the error for the AP scheme. However, for large $\alpha$ both errors are close to the error of the implicit scheme that can be regarded as a reference solution for fixed $\dx$.
\begin{table}[ht!]
 \begin{center}
    \begin{tabular}{|c||c||c|c|c|c|}
    \hline$\alpha^{-2}$ & $\dx $  				   &   0.01 &  0.005 & 0.0025 & 0.00125\\\hline \hline
         10& CPU-time [s]						   &   15.9 &   32.1 &   89.1 &  471.0 \\
     	   &$\|\bar{\vecu}-\vecu^\alpha_h\|_{L^2}$ & 5.4e-1 & 6.3e-2 & 1.3e-2 & 3.4e-3 \\\hline
        100& CPU-time [s]						   &   15.8 &   32.1 &   89.3 &  470.8 \\
     	   &$\|\bar{\vecu}-\vecu^\alpha_h\|_{L^2}$ & 5.4e-1 & 5.1e-2 & 1.2e-2 & 3.3e-3 \\\hline
       1000& CPU-time [s]						   &   16.2 &   32.9 &   88.0 &  466.4 \\
     	   &$\|\bar{\vecu}-\vecu^\alpha_h\|_{L^2}$ & 4.7e-1 & 5.0e-2 & 1.2e-2 & 2.5e-3 \\\hline
      10000& CPU-time [s]						   &   15.7 &   32.5 &   89.3 &  467.5 \\
     	   &$\|\bar{\vecu}-\vecu^\alpha_h\|_{L^2}$ & 4.6e-1 & 5.0e-2 & 1.1e-2 & 2.4e-3 \\\hline
     100000& CPU-time [s]						   &   15.7 &   33.2 &   88.2 &  468.0 \\
     	   &$\|\bar{\vecu}-\vecu^\alpha_h\|_{L^2}$ & 4.6e-1 & 5.0e-2 & 1.1e-2 & 2.2e-3 \\\hline
    \end{tabular}
 \caption{CPU-times and discrete $L_2$-distances between the solution of the AP scheme and the reference solution $\bar{\vecu}$. The reference solution was computed with an implicit DG scheme.}   
 \label{tab:IMEXerror}
  \end{center}
 \end{table}
 \begin{table}[ht!]
 \begin{center}
   \begin{tabular}{|c||c||c|c|c|c|}
       \hline$\alpha^{-2}$ & $\dx$       		 &   0.01 &  0.005 & 0.0025 & 0.00125\\\hline \hline
        10    &CPU-time [s]                      &    50 &     103 &    229 &    483 \\
        &$\|\bar{\vecu} -\vecu^\alpha_h\|_{L^2}$ & 7.7e-2 & 5.1e-2 & 1.2e-2 & 3.6e-3 \\\hline
        100   &CPU-time [s]                      &    133 &    440 &    615 &   1403 \\
        &$\|\bar{\vecu} -\vecu^\alpha_h\|_{L^2}$ & 7.3e-2 & 5.1e-2 & 8.8e-3 & 3.0e-3 \\\hline
        1000  &CPU-time [s]                      &    440 &    857 &   1736 &   3855 \\
        &$\|\bar{\vecu} -\vecu^\alpha_h\|_{L^2}$ & 7.0e-2 & 3.3e-2 & 8.5e-3 & 2.5e-3 \\\hline
        10000 &CPU-time [s]                      &   1117 &   2375 &   5687 &  13154 \\
        &$\|\bar{\vecu} -\vecu^\alpha_h\|_{L^2}$ & 6.9e-2 & 3.2e-2 & 8.3e-3 & 2.2e-3 \\\hline
        100000&CPU-time [s]                      &   3694 &   7949 &  16246 &  39526 \\
        &$\|\bar{\vecu} -\vecu^\alpha_h\|_{L^2}$ & 6.9e-2 & 3.1e-2 & 8.2e-3 & 2.1e-3 \\\hline\hline
\rule{0pt}{12pt}NSK &$\|\bar{\vecu} -\vecu^\text{NSK}_h\|_{L^2}$ & 6.8e-2 & 3.0e-2 & 8.0e-3 & 2.0e-3 \\\hline
    \end{tabular}
     \caption{CPU-times and discrete $L_2$-distances between the solution of the discontinuous Galerkin scheme and the reference solution $\bar{\vecu}$. The last line contains the discrete $L_2$-distance for an explicit DG discretization of the NSK system. The reference solution was computed with an implicit DG scheme.} 
     \label{tab:DGerror}
 \end{center}
 \end{table}

In summary, we observe that the AP scheme \eqref{eq:slowd}, \eqref{eq:fastd} is able to provide solutions with the same magnitude of error as the DG scheme much faster. The AP scheme provides the numerical solution independent of the parameter $\alpha$ and thus is more efficient for small $\alpha$.


\subsection{Large density variations}
We already pointed out in Section \ref{sec:intro} that explicit numerical methods for the NSK system \eqref{eq:NSK} are not able to deal with large density jumps. As we explained above the first-order part of the system is of mixed hyperbolic-elliptic type and one cannot use shock-capturing numerical solvers for computations. Thus, there is no chance to stabilize the numerical scheme for large density gradients. In \cite{NeusserRohdeSchleper-15}, it was shown that this problem could be overcome if one uses the relaxation system \eqref{eq:aNSK}.

In this 1-D test case, we introduce a modified version of the Van-der-Waals pressure equation \eqref{eq:vdw}
\[
p^s(\rho)=s\cdot p\left(\frac{\rho}{s}\right),
\]
with the scaling parameter $s=100$ to enlarge the elliptic region (cf. \cite{Blesgen}). This enables large density jumps for phase boundaries. We apply the AP scheme \eqref{eq:slowd}, \eqref{eq:fastd} to the relaxation system \eqref{eq:aNSK}, which is considered with $\gamma=0.16,~\gamma\eps^2=10^{-5}$ and subject to the following initial conditions:
\begin{equation}
\begin{aligned}
\rho_0(x)&=\left\{\begin{aligned}
&30: &&x\in
\begin{array}{l}
(0.08,0.12)\cup(0.14,0.16)\cup(0.18,0.22)\cup(0.30,0.32)\cup\\
(0.38,0.42)\cup(0.44,0.46)\cup(0.48,0.52)\cup(0.58,0.62)\cup\\
(0.70,0.72)\cup(0.78,0.82)\cup(0.84,0.86)\cup(0.88,0.92)
\end{array}
\\
&180: &&\mbox{otherwise},\end{aligned}\right.\\
v_0(x)&=0.
\end{aligned}
\label{eq:initialbubble1Dlarge}
\end{equation}

\begin{figure}[ht!]
\subfigure[t=0]{\includegraphics[width=0.32\textwidth]{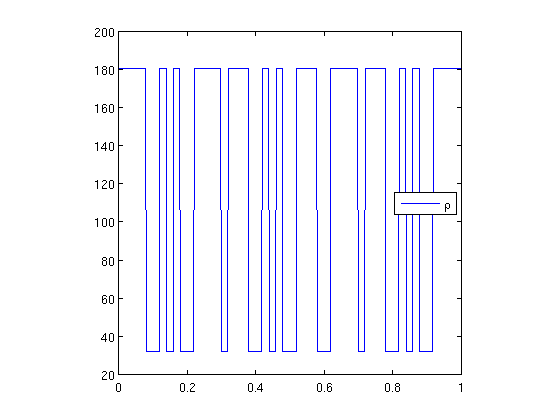}}
\subfigure[t=0.5]{\includegraphics[width=0.32\textwidth]{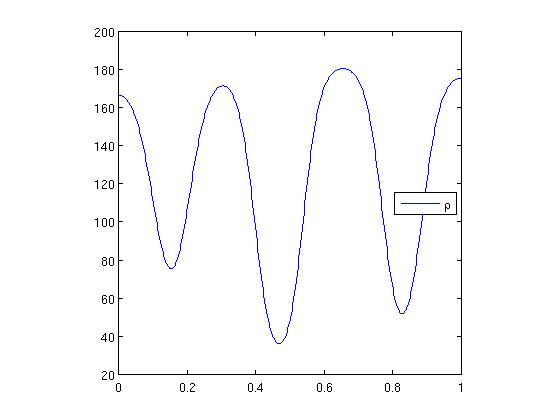}} 
\subfigure[t=1]{\includegraphics[width=0.32\textwidth]{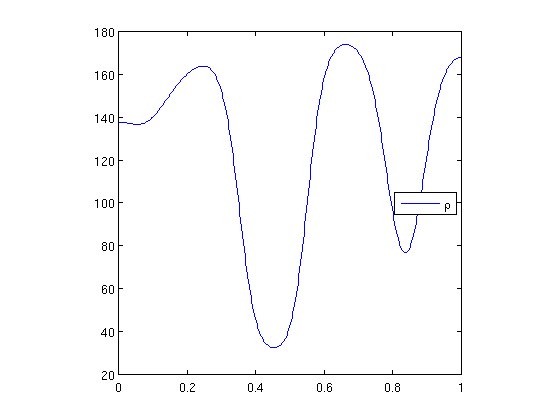}}\\
\subfigure[t=2]{\includegraphics[width=0.32\textwidth]{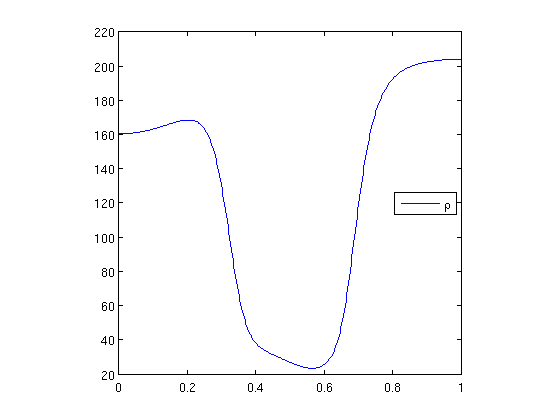}}
\subfigure[t=5]{\includegraphics[width=0.32\textwidth]{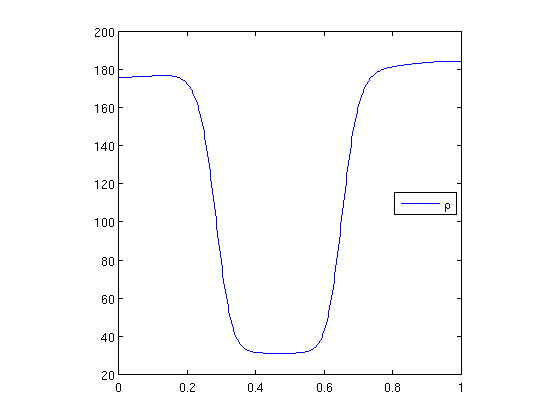}}
\subfigure[t=15]{\includegraphics[width=0.32\textwidth]{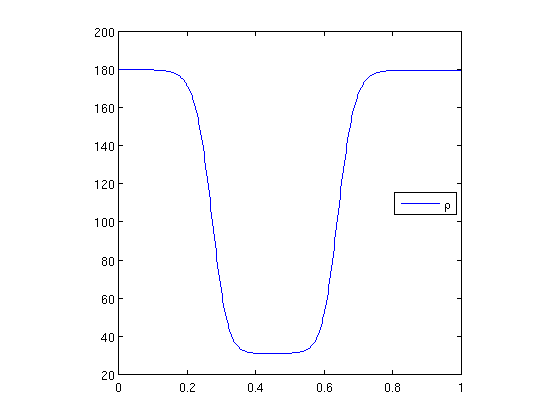}}
\caption{Density  evolution for  the relaxation system \rf{eq:NSK_a} with $\dx=0.00125$, $\gamma\eps^2=1e-5$, $\alpha=100000$ and initial datum \rf{eq:initialbubble1Dlarge}. The small bubbles shrink and merge (t=0.5, t=1) and form one bubble (t=2). At t=15 the material seems to be in equilibrium.}
\label{fig:large}
\end{figure}

\begin{table}[ht!]
 \begin{center}
    \begin{tabular}{c||c|c|c|c|c|c}
    			i &	      1 & 	  2 &	    3 &	     4 &	      5 &	  6\\\hline	
   	$\alpha^{-2}$ &      10 &     100 &    1000 &   10000 &  100000 & 1000000\\\hline\hline
 \rule{0pt}{12pt}$\|\vecu^\text{NSK}_h -\vecu^\alpha_h\|_{L^2}$
 				  & 2.21e-1 & 3.15e-2 & 3.30e-3 & 3.31e-4 & 3.30e-5 & 4.07e-6\\\hline
 \rule{0pt}{15pt}EOC$_i=\frac{{\rm ln}(D_h^i/D_h^{i+1})}{{\rm ln}(\alpha^2_{i}/\alpha^2_{i+1})}$
     			  &       - &    0.84 &    0.98 &    1.00 &    1.00 &   0.90\\\hline
     CPU-time [s] &     594 &     594 &     594 &     596 &     594 &    594
    \end{tabular}
 \end{center}
  \caption{Discrete $L_2(\Omega)$-distance for $\dx=0.00125$ between the solution of the AP scheme and a reference solution that was computed with an implicit DG scheme. The distance decreases as $\alpha$ does. The second line contains the  experimental order of convergence (EOC) with respect to $\alpha$ and the third line contains the CPU time.}
    \label{tab:large}
\end{table}

Table \ref{tab:large} and Figure \ref{fig:large} indicate that it is possible to simulate large density rations with the AP scheme \eqref{eq:slowd}, \eqref{eq:fastd}. 
The first and second line of Table \ref{tab:large} show that the scheme is asymptotic preserving in this test case, as the discrete $L_2$-distance decreases as $\alpha$ does. The third line underlines, that, analogously to the previous test case, the CPU time does not depend on the Korteweg parameter $\alpha$.


\subsection{Sharp interface limit}
We again consider the 1-D version of the relaxation system \eqref{eq:aNSK} with $\alpha=100000,~\gamma=0.16$ and the   following initial conditions:
\begin{equation}
\begin{array}{lll}
\rho_0(x)&=\left\{\begin{array}{ll}0.3: &x\in(0.3,0.7)\\
1.8: &\mbox{otherwise}\end{array}\right.,\\
v_0(x)&=0.
\end{array}
\label{eq:initialsharpBubble1D}
\end{equation}
The initial datum is a non-equilibrium bubble that will be driven towards a two-phase equilibrium by the evolution of the relaxation system. We want to study if the AP scheme \eqref{eq:slowd}, \eqref{eq:fastd} can deal with tiny interfaces which appear in the sharp interface limit ($\eps\to0$) for fixed $\dx=0.00125$.

\begin{figure}[ht!]
\subfigure[]{\includegraphics[width=0.5\textwidth]{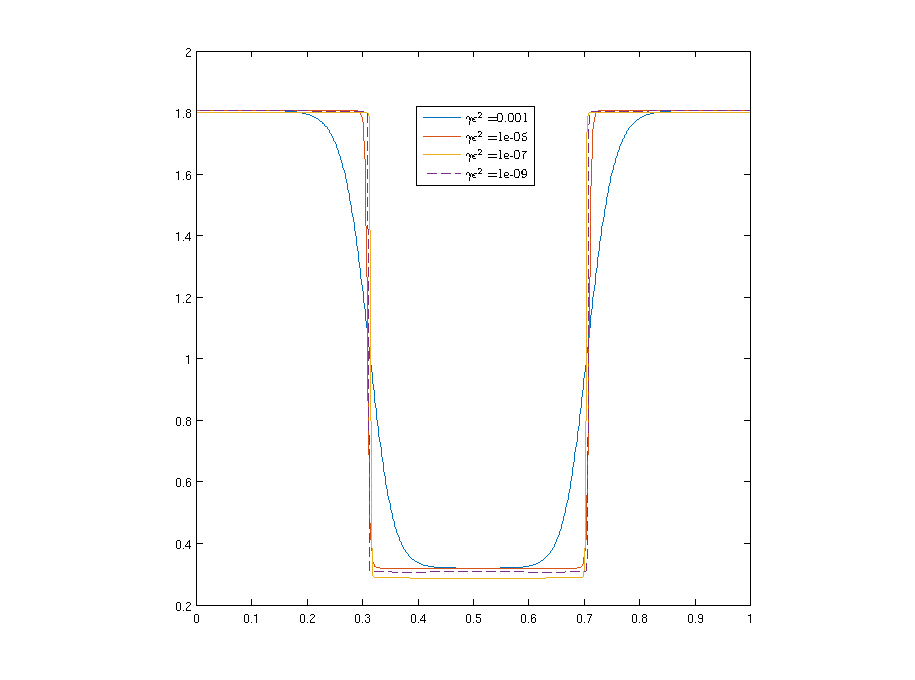}}
\subfigure[Zoom]{\includegraphics[width=0.5\textwidth]{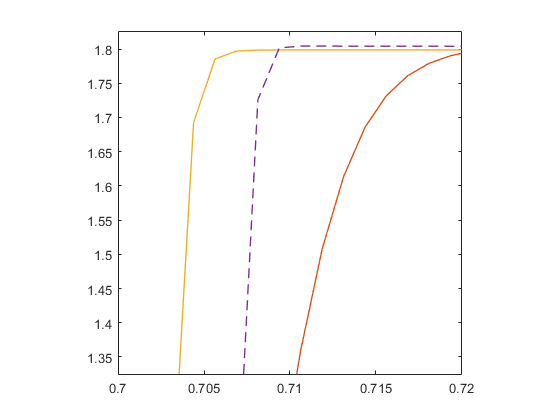}} 
\caption{Approximate density distribution at $t=15$ for different $\eps$. The right Figure displays a zoom of the upper corner of the right interface}
\label{fig:sharpBubble1D}
\end{figure}

The density distributions computed by AP scheme \eqref{eq:slowd}, \eqref{eq:fastd} with $\dx=0.00125$ are plotted Figure \ref{fig:sharpBubble1D} doe different values of $\eps$. As one can see, shows the AP scheme is able to accurately capture tiny interfaces which appear in the sharp interface limit ($\eps\to0$). This means that we are able to obtain numerical solutions that show the expected behavior even for underresolved meshes.
However, we were not able to give analytical proof for this property of the scheme.

\begin{table}[ht!]
 \begin{center}
    \begin{tabular}{c||c|c|c|c|c|c|c}	
   	~$\gamma\eps^2$ = &$10^{-3}$&$10^{-4}$&$10^{-5}$&$10^{-6}$&$10^{-7}$&$10^{-8}$&$10^{-9}$\\\hline\hline
    CPU-time [s] & 8001 & 2599 &  831 &  465 &  463 &  464 &  461
    \end{tabular}
 \end{center}
  \caption{CPU time for different $\eps$. The CPU-time decreases as $\eps$ does.}
    \label{tab:sharp}
\end{table}
Another property of the AP scheme \eqref{eq:slowd}, \eqref{eq:fastd} can be seen in Table \ref{tab:sharp}. For decreasing values of $\eps$ the CPU time decreases. For $\gamma\eps^2\leq 10^{5}$ the CPU-times are constant. We recall \rf{eq:cfl} and see that the diffusive term dominates for larger $\eps$ while the convective term does so for smaller $\eps$. Additionally, the solution of the fast system is very time consuming for the small time steps that occur for larger $\eps$. In fact, it is a nice feature of the AP scheme, that it is faster for smaller and thus more realistic $\eps$. 


\subsection{2D test case: static equilibrium}
In this example, we compute a numerical solution to the relaxation system \rf{eq:aNSK} in the 2-D case. We start with 22 bubbles in a bounded box $(0,1)^2$ filled with liquid. We choose $\alpha^{-2}=1000,~\gamma=1,~\eps=0.01$. 
\begin{figure}[ht!]
\subfigure[t=0   ]{\includegraphics[width=0.37\textwidth]{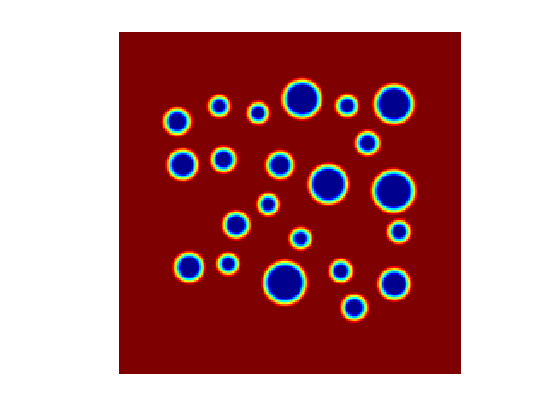}}\hspace{-.9cm}
\subfigure[t=0.06]{\includegraphics[width=0.37\textwidth]{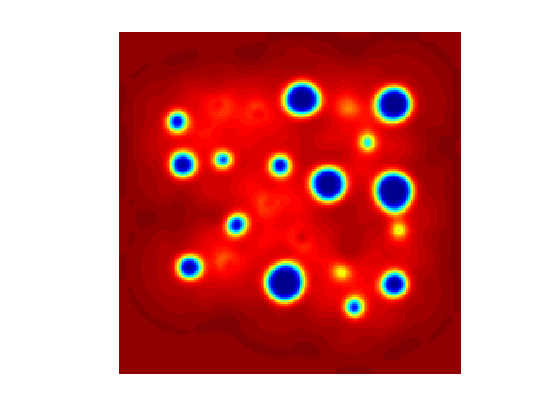}}\hspace{-.9cm} 
\subfigure[t=0.1 ]{\includegraphics[width=0.37\textwidth]{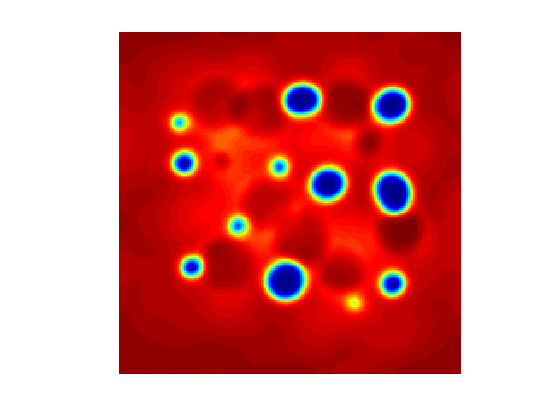}}\\
\subfigure[t=1   ]{\includegraphics[width=0.37\textwidth]{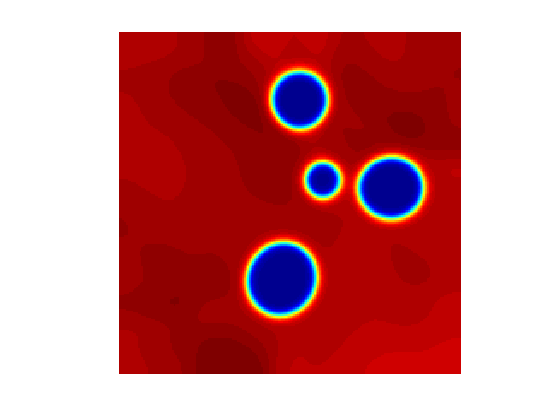}}\hspace{-.9cm}
\subfigure[t=3   ]{\includegraphics[width=0.37\textwidth]{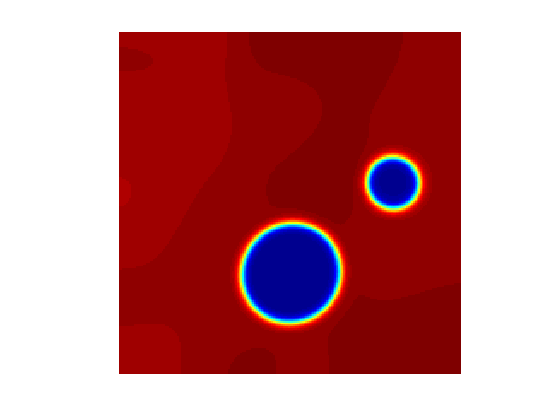}}\hspace{-.9cm}
\subfigure[t=5   ]{\includegraphics[width=0.37\textwidth]{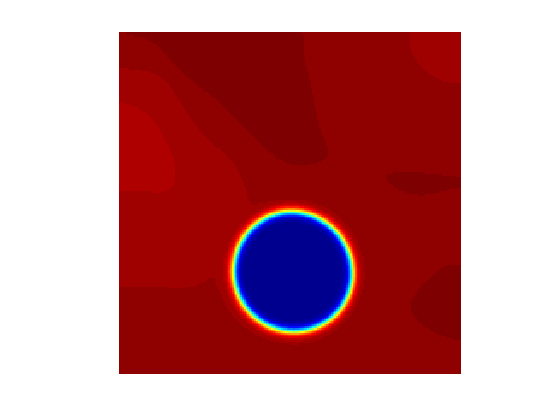}}
\caption{Density evolution for the relaxation system \rf{eq:NSK_a} and $\dx=\dy=0.005$.  }
\label{fig:Bubbles2Dcoarse}
\end{figure}
\begin{figure}[ht!]
\subfigure[t=0   ]{\includegraphics[width=0.37\textwidth]{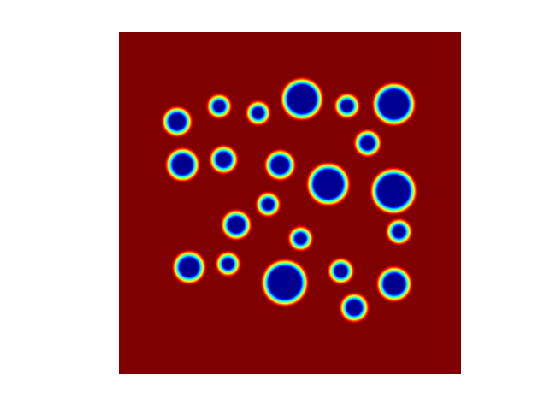}}\hspace{-.9cm}
\subfigure[t=0.06]{\includegraphics[width=0.37\textwidth]{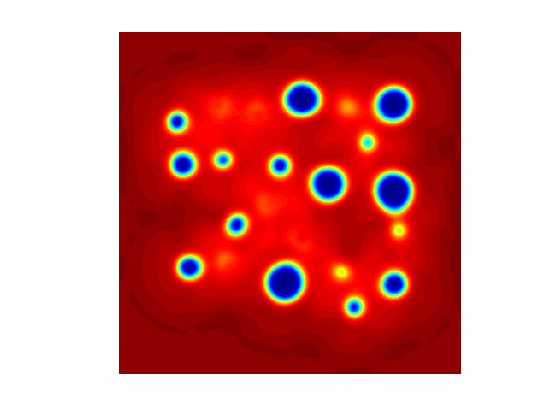}}\hspace{-.9cm} 
\subfigure[t=0.1 ]{\includegraphics[width=0.37\textwidth]{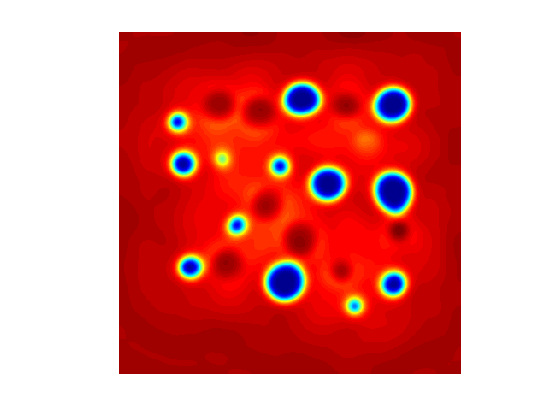}}\\
\subfigure[t=1   ]{\includegraphics[width=0.37\textwidth]{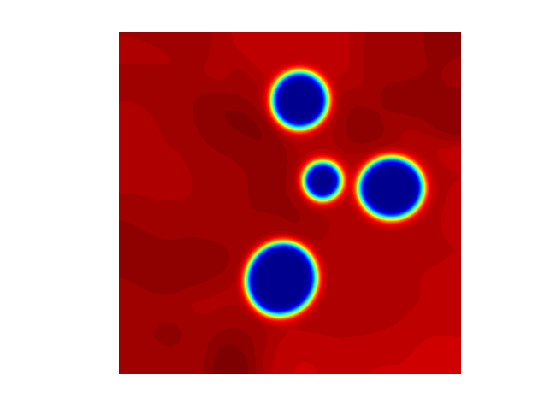}}\hspace{-.9cm}
\subfigure[t=3   ]{\includegraphics[width=0.37\textwidth]{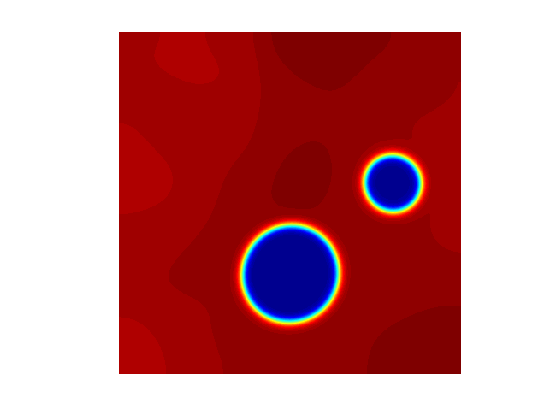}}\hspace{-.9cm}
\subfigure[t=5   ]{\includegraphics[width=0.37\textwidth]{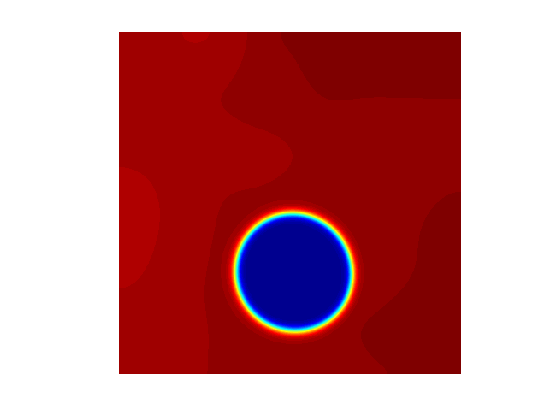}}
\caption{Density evolution for the relaxation system \rf{eq:NSK_a} and $\dx=\dy=0.0025$.  }
\label{fig:Bubbles2D}
\end{figure}

We use a uniform rectangular grid and run the simulation up to $t=5$ on two different grids with $\dx=\dy=0.005$ and $\dx=\dy=0.0025$. The results (the density distribution) computed at different times are plotted in 
Figures \ref{fig:Bubbles2Dcoarse} and \ref{fig:Bubbles2D}. The density varies between 0.3 (blue) and 1.8 (red). At $t=0.06,~t=0.1$ bubbles collapse and emit shock waves. At $t=1,~t=3$ the smaller bubbles shrink and the biggest bubble grows. At $t=5$ the fluid seems to be in an equilibrium configuration. Note that the position of the bubbles and the coalescence dynamics are the same for both gird widths. Even though we are not able to resolve the interface by the underlying grid due to CPU time constraints, the results show the desired behavior for a two-phase system.  The computations for different grid sizes support the statement, that the behavior of the solution is not only qualitatively but also quantitatively correct. However, to be absolutely sure, one has to compare it with a fully resolved solution. In order to run such a simulation, one would need to implement a scheme that incorporates adaptive time and spatial stepping, which is out of the scope of the current paper.


\section{Conclusion/Outlook}
In this work, we presented a new asymptotic-preserving scheme for a relaxation of the Navier-Stokes-Korteweg system. The method has two key ingredients. First, we introduced a modified pressure that guarantees that the Euler part of the system is hyperbolic. Second, we split the relaxation system into a non linear conservative system for the slow dynamics and a non conservative stiff system for the fast dynamics that come from the Korteweg term. 
We showed that our scheme is asymptotic preserving, i.e. it converges to a numerical  scheme for the Navier-Stokes-Korteweg equations on the fully discrete level. 
We supported this result with several numerical tests. In all test cases, the solutions showed the expected behavior. We were able to compute large density ratios and small interfacial widths for fixed grid widths and we were able to obtain approximate solutions of the same quality as those of an explicit scheme, but with significantly smaller computational costs. 
Additionally, our scheme showed to be asymptotic preserving in the sharp interface limit in our numerical tests. Note that up to now we have not been able to give analytical proof to that. 

The numerical proof of the asymptotic preserving property is one goal for future work. This is by no means an easy task, because the system has a jump of the density at the phase boundaries. Therefore one needs to impose special coupling conditions at the interface and to introduce an interface tracking algorithm. 
Another main goal is the extension of the numerical algorithm to three space dimensions. In order to accomplish this,  we will look to introduce an adaptive time and spatial stepping and to use a discontinuous Galerkin formulation instead of the finite volume formulation. These two extensions guarantee that the solution of the linear system remains efficient, because we are able to control the number of cells, even if we need to refine the grid at the interface. This could prove to be important for the simulation of realistic two-phase flow examples such as underwater explosions. 
\section*{Acknowledgment}
This material is based upon work supported by the NSF
Research Network Grant no. RNMS11-07444 (KI-Net). The work of AC was supported in part by the NSF grants DMS-1216974 and DMS-1521051. PD acknowledges
support from EPSRC under grant no. EP/M006883/1 and from the Royal Society
and the Wolfson Foundation through a Royal Society Wolfson Research
Merit Award. PD is on leave from CNRS, Institut de Math\'ematiques de
Toulouse, France. AC and JN gratefully acknowledge the hospitality of the
Department of Mathematics, Imperial College London, where part of this 
research has been conducted.  JN acknowledges the support by the German Research Foundation
(DFG) in the framework of the Collaborative Research Center Transregio 75 Droplet Dynamics
under Extreme Ambient Conditions and by the Elite program for postDocs of the Baden-W\"urttemberg
Foundation.
\newpage
\bibliographystyle{plain} 
%

\begin{thebibliography}{10}

\bibitem{ADGK13}
G.~L. Aki, W.~Dreyer, J.~Giesselmann, and C.~Kraus.
\newblock {A quasi-incompressible diffuse interface model with phase
  transition}.
\newblock {\em Mathematical Models and Methods in Applied Sciences},
  24(5):827--861, 2014.

\bibitem{AMW-98}
D.~M. Anderson, G.~B. McFadden, and A.~A. Wheeler.
\newblock Diffuse-interface methods in fluid mechanics.
\newblock In {\em Annual review of fluid mechanics}, volume~30, pages 139--165.
  Annual Reviews, Palo Alto, CA, 1998.

\bibitem{BassiRebay}
F.~Bassi and S.~Rebay.
\newblock A high-order accurate discontinuous finite element method for the
  numerical solution of the compressible {N}avier-{S}tokes equations.
\newblock {\em Journal of Computational Physics}, 131(2):267--279, 1997.

\bibitem{Benzoni99}
S.~Benzoni-Gavage.
\newblock {Nonuniqueness of phase transitions near the {M}axwell line}.
\newblock {\em Proceedings of the American Mathematical Society},
  127(4):1183--1190, 1999.

\bibitem{Blesgen}
T.~{Blesgen}.
\newblock {A generalization of the Navier-Stokes equations to two-phase flows}.
\newblock {\em Journal of Physics D Applied Physics}, 32:1119--1123, May 1999.

\bibitem{BraackProhl}
M.~Braack and A.~Prohl.
\newblock {Stable discretization of a diffuse interface model for liquid-vapor
  flows with surface tension}.
\newblock {\em ESAIM. Mathematical Modelling and Numerical Analysis},
  47(2):401--420, 2013.

\bibitem{Bresch-03}
D.~Bresch, B.~Desjardins, and C.-K. Lin.
\newblock {On some compressible fluid models: {K}orteweg, lubrication, and
  shallow water systems}.
\newblock {\em Communications in Partial Differential Equations},
  28(3-4):843--868, 2003.

\bibitem{Charve13}
F.~{Charve}.
\newblock {Convergence of a low order non-local {N}avier-{S}tokes-{K}orteweg
  system: the order-parameter model}.
\newblock {\em http://arxiv.org/abs/1302.2617}, February 2013.

\bibitem{Charve-13}
F.~Charve.
\newblock {Local in time results for local and non-local capillary
  {N}avier--{S}tokes systems with large data}.
\newblock {\em Journal of Differential Equations}, 256(7):2152--2193, 2014.

\bibitem{CockburnShu-III}
B.~Cockburn, S.~Y. Lin, and C.-W. Shu.
\newblock {T{VB} {R}unge-{K}utta local projection discontinuous {G}alerkin
  finite element method for conservation laws. {III}. {O}ne-dimensional
  systems}.
\newblock {\em Journal of Computational Physics}, 84(1):90--113, 1989.

\bibitem{CGRS}
F.~Coquel, E.~Godlewski, P.-A. Raviart, and N.~Seguin.
\newblock An asymptotic preserving scheme for {E}uler equations with gravity
  and friction.
\newblock In {\em Finite volumes for complex applications {V}}, pages 305--312.
  ISTE, London, 2008.

\bibitem{CDK}
F.~Cordier, P.~Degond, and A.~Kumbaro.
\newblock An asymptotic-preserving all-speed scheme for the {E}uler and
  {N}avier-{S}tokes equations.
\newblock {\em J. Comput. Phys.}, 231(17):5685--5704, 2012.

\bibitem{Corli-Rohde-12}
A.~Corli and C.Rohde.
\newblock {Singular limits for a parabolic--elliptic regularization of scalar
  conservation laws}.
\newblock {\em Journal of Differential Equations}, 253(5):1399--1421, 2012.

\bibitem{DegondJinLiu-07}
P.~Degond, S.~Jin, and J.~G. Liu.
\newblock {Mach-number uniform asymptotic-preserving gauge schemes for
  compressible flows }.
\newblock {\em Bulletin of the Institute of Mathematics, Academia Sinica, New
  Series}, 2(4):851--892, 2007.

\bibitem{DT11}
P.~Degond and M.~Tang.
\newblock All speed scheme for the low {M}ach number limit of the isentropic
  {E}uler equations.
\newblock {\em Commun. Comput. Phys.}, 10(1):1--31, 2011.

\bibitem{Diehl}
D.~Diehl.
\newblock {\em Higher Order Schemes for Simulation of Compressible Liquid-Vapor
  Flows with Phase Change}.
\newblock PhD thesis, {A}lbert-{L}udwigs-{U}niversit{\"a}t {F}reiburg, 2007.

\bibitem{DreyerGiesselmannKrausRohde}
W.~Dreyer, J.~Giesselmann, C.~Kraus, and C.~Rohde.
\newblock {Asymptotic analysis for {K}orteweg models}.
\newblock {\em Interfaces and Free Boundaries. Mathematical Modelling, Analysis
  and Computation}, 14(1):105--143, 2012.

\bibitem{DunnSerrin-85}
J.E. Dunn and J.~Serrin.
\newblock {On the thermomechanics of interstitial working}.
\newblock {\em Archive for Rational Mechanics and Analysis}, 88(2):95--133,
  1985.

\bibitem{Giesselmann14a}
J.~Giesselmann.
\newblock {A Relative Entropy Approach to Convergence of a Low Order
  Approximation to a Nonlinear Elasticity Model with Viscosity and
  Capillarity}.
\newblock {\em SIAM Journal on Mathematical Analysis}, 46(5):3518--3539, 2014.

\bibitem{Giesselmann-15}
J.~Giesselmann.
\newblock {Low {M}ach asymptotic-preserving scheme for the {E}uler-{K}orteweg
  model}.
\newblock {\em IMA Journal of Numerical Analysis}, 35(2):802--833, 2014.

\bibitem{GLT15}
J.~Giesselmann, C.~Lattanzio, and A.~Tzavaras.
\newblock Relative energy for the {K}orteweg theory and related {H}amiltonian
  flows in gas dynamics.
\newblock 2015.
\newblock http://arxiv.org/abs/1510.00801.

\bibitem{Jinetal-99}
F.~Golse, S.~Jin, and C.~D. Levermore.
\newblock {The Convergence of Numerical Transfer Schemes in Diffusive Regimes
  I: Discrete-Ordinate Method}.
\newblock {\em SIAM Journal on Numerical Analysis}, 36(5):1333--1369, 1999.

\bibitem{Haack-12}
J.~Haack, S.~Jin, and J.-G. Liu.
\newblock {An all-speed asymptotic-preserving method for the isentropic {E}uler
  and {N}avier-{S}tokes equations}.
\newblock {\em Communications in Computational Physics}, 12(4):955--980, 2012.

\bibitem{HattoriLi-94}
H.~Hattori and D.~Li.
\newblock {Solutions for two-dimensional system for materials of {K}orteweg
  type}.
\newblock {\em SIAM Journal on Mathematical Analysis}, 25(1):85--98, 1994.

\bibitem{Hermsdorferetal}
K.~Hermsd{\"o}rfer, C.~Kraus, and D.~Kr{\"o}ner.
\newblock {Interface conditions for limits of the {N}avier-{S}tokes-{K}orteweg
  model}.
\newblock {\em Interfaces and Free Boundaries. Mathematical Modelling, Analysis
  and Computation}, 13(2):239--254, 2011.

\bibitem{Jacobs-95}
D.~Jacobs, B.~Mckinney, and M.~Shearer.
\newblock {Traveling Wave Solutions of the Modified Korteweg-deVries-Burgers
  Equation}.
\newblock {\em Journal of Differential Equations}, 116(2):448--467, 1995.

\bibitem{Jamet-01}
D.~Jamet, O.~Lebaigue, N.~Coutris, and J.M. Delhaye.
\newblock The second gradient method for the direct numerical simulation of
  liquid--vapor flows with phase change.
\newblock {\em Journal of Computational Physics}, 169(2):624--651, 2001.

\bibitem{Jin-99}
S.~Jin.
\newblock {Efficient Asymptotic-Preserving (AP) schemes for some multiscale
  kinetic wquations}.
\newblock {\em SIAM Journal on Scientific Computing}, 21(2):441--454, 1999.

\bibitem{JinLevermore-91}
S.~Jin and D.~Levermore.
\newblock {The discrete-ordinate method in diffusive regimes}.
\newblock {\em Transport Theory and Statistical Physics}, 20(5-6):413--439,
  1991.

\bibitem{Kle95}
R.~Klein.
\newblock Semi-implicit extension of a {G}odunov-type scheme based on low
  {M}ach number asymptotics. {I}. {O}ne-dimensional flow.
\newblock {\em J. Comput. Phys.}, 121(2):213--237, 1995.

\bibitem{Korteweg-01}
D.~J. Korteweg.
\newblock {Sur la forme que prennent les {\'e}quations du mouvement des fluides
  si l'on tient compte des forces capillaires caus{\'e}es par des variations de
  densit{\`e} consid{\'e}rables mais continues et sur la th{\`e}orie de la
  capillarit{\'e} dans l'hypoth{\`e}se d'une variation continue de la
  densit{\'e}}.
\newblock {\em Archives N{\'e}erlandaises de Sciences Exactes et Naturelles},
  2(6):1--24, 1901.

\bibitem{Kotschote-08}
M.~Kotschote.
\newblock {Strong solutions for a compressible fluid model of {K}orteweg type}.
\newblock {\em Annales de l'Institut Henri Poincar{\'e} (C) Non Linear
  Analysis}, 25(4):679--696, 2008.

\bibitem{Larsen-87}
E.~W Larsen, J.~E. Morel, and W.~F. Miller~Jr.
\newblock {Asymptotic solutions of numerical transport problems in optically
  thick, diffusive regimes}.
\newblock {\em ~Journal of Computational Physics}, 69(2):283--324, 1987.

\bibitem{NeusserRohdeSchleper-15}
J.~Neusser, C.~Rohde, and V.~Schleper.
\newblock Relaxation of the {N}avier-{S}tokes-{K}orteweg equations for
  compressible two-phase flow with phase transition.
\newblock {\em International Journal for Numerical Methods in Fluids}, 2015.

\bibitem{ANLM}
S.~Noelle, G.~Bispen, K.~R. Arun,
  M.~Luk{\'a}{\v{c}}ov{\'a}-Medvi{\v{d}}ov{\'a}, and C.-D. Munz.
\newblock A weakly asymptotic preserving low {M}ach number scheme for the
  {E}uler equations of gas dynamics.
\newblock {\em SIAM J. Sci. Comput.}, 36(6):B989--B1024, 2014.

\bibitem{PR03}
L.~Pareschi and G.~Russo.
\newblock High order asymptotically strong-stability-preserving methods for
  hyperbolic systems with stiff relaxation.
\newblock In {\em Hyperbolic problems: theory, numerics, applications}, pages
  241--251. Springer, Berlin, 2003.

\bibitem{Rohde-10}
Christian Rohde.
\newblock A local and low-order {N}avier-{S}tokes-{K}orteweg system.
\newblock In {\em Nonlinear partial differential equations and hyperbolic wave
  phenomena}, volume 526, pages 315--337. Amer. Math. Soc., 2010.

\bibitem{Solci-02}
Margherita Solci.
\newblock {\em {A Variational Model for Phase Separation}}.
\newblock PhD thesis, Scuola {N}ormale {S}uperiore {P}isa, 2002.

\bibitem{Tadmor-84}
E.~Tadmor.
\newblock {Numerical viscosity and the entropy condition for conservative
  difference schemes}.
\newblock {\em Mathematics of Computation}, 43(168):369--381, 1984.

\bibitem{Torobook}
E.~F. Toro.
\newblock {\em Riemann solvers and numerical methods for fluid dynamics}.
\newblock Springer-Verlag, Berlin, third edition, 2009.
\newblock A practical introduction.

\bibitem{Truskinovsky-94}
L.~Truskinovsky.
\newblock {About the ``normal growth'' approximation in the dynamical theory of
  phase transitions}.
\newblock {\em Continuum Mechanics and Thermodynamics}, 6(3):185--208, 1994.

\bibitem{Witterstein}
G.~Witterstein.
\newblock {Phase change flows with mass exchange}.
\newblock {\em Advances in Mathematical Sciences and Applications},
  21(2):559--611, 2011.

\end{thebibliography}

\end{document}